\documentclass[prd,nofootinbib,superscriptaddress,twocolumn,preprintnumbers]{revtex4-1}
\makeatletter
\def\l@subsubsection#1#2{}
\makeatother
\usepackage{placeins}

\usepackage{xcolor}
\usepackage{booktabs}
\usepackage{amsmath,epsfig}
\allowdisplaybreaks[1]

\usepackage{pifont}
\usepackage[utf8]{inputenc}

\usepackage[
 final,
 colorlinks=true
,urlcolor=blue
,anchorcolor=blue
,citecolor=blue
,filecolor=blue
,linkcolor=blue
,menucolor=blue
,linktocpage=true
,pdfproducer=medialab
,pdfa=true
]{hyperref}
\usepackage{natbib}
\usepackage{hypernat}

\newcommand{\mel}[3]{\langle #1 | #2 | #3\rangle} % Matrix Element
\newcommand{\GeV}{\,\mathrm{GeV}}

\newcommand{\eVunit}{\mathrm{eV}}

\begin{document}

\preprint{CERN-TH-2025-246}

\title{Probing invisible particles with charm}

\author{Gudrun Hiller}
\email{gudrun.hiller@cern.ch}
\affiliation{TU Dortmund University, Department of Physics, Otto-Hahn-Str.4, D-44221 Dortmund, Germany}
\affiliation{Theoretical Physics Department, CERN, 1211 Geneva 23, Switzerland}
\author{Dominik Suelmann}
\email{dominik.suelmann@tu-dortmund.de}
\affiliation{TU Dortmund University, Department of Physics, Otto-Hahn-Str.4, D-44221 Dortmund, Germany}

\begin{abstract}
We point out  opportunities to probe  invisible particles,  left- and right-handed neutrinos,  axion-like particles (ALPs) and dark photons $(Z^\prime)$  with rare 
decays of charm hadrons. We employ and recast existing searches in $D \to (\pi, \omega) X$, $D^ 0 \to X$
and $\Lambda_c \to p X$, where $X$ denotes one of the above  invisible final states including dineutrinos. The branching ratios are clean  null tests
of the standard model, yet, are essentially unconstrained for some parameters  of  light new physics, limited only by weak lifetime constraints at the level of $\mathcal{O}(10^{-1})$.
On the other hand, if models are probed, branching ratios still reach up to $10^{-3}$ ($Z^\prime$) and  $10^{-4}$ (ALPs).
Chirality-preserving operators from heavy new physics in the dimension six standard model effective theory (SMEFT) imply tighter upper limits, up to few $\times 10^{-5}$. 
Constraints on chirality-flipping heavy new physics, such as  lepton number violation  from dimension seven SMEFT, or 
with light sterile neutrinos,
are weaker,  with branching ratios up to few$\times 10^{-4}$. Sensitivities to different couplings  arise with  $\Lambda_c \to p X $  and $D \to \pi \pi X$ decays, in particular in relation with the other modes.
Processes can be studied at  running  and future experiments with high charm luminosities, BESIII, Belle II, a  super-tau-charm factory  (STCF) and
$Z$-factories, such as the FCC-ee and the CEPC. 
\end{abstract}

\maketitle

\tableofcontents

\section{Introduction}

Missing energy signatures allow to   explore  different  physics  scenarios in one go:
they are sensitive to neutrino interactions, including  sterile ones,  as well as phenomena such as lepton number violation (LNV).
Missing energy can also stem from models beyond the standard model, with  long-lived new particles  decaying  outside of  detectors, or into a dark, invisible  sector, opening the way  to probe the latter.
Looking for  invisibles  in rare, flavor changing neutral current (FCNC) quark transitions combines the best of both worlds, predictivity and sensitivity, and complements precision flavor studies with leptonic and radiative final states.

In this work we consider rare, FCNC  processes in the up-quark sector, of charm quarks into  invisible final states.
Charm physics opens a unique window into  the flavor landscape, complementing existing precision programs with kaons  and $b$-hadrons.
A genuine feature of $|\Delta c|=|\Delta u|=1$ transitions is an efficient GIM-suppression, which in particular renders $c \to u \nu \bar \nu$ transitions in the standard model (SM) too small to be observable  in the foreseeable future~\cite{Burdman:2001tf}. This identifies
the dineutrino modes, or more general, decays into invisible final states,  as clean null tests  of the SM.

To date, only  two  FCNC modes into invisibles have been measured, $\mathcal{B}(K^+ \to \pi^+ \nu \bar \nu)=(1.14^{+0.40}_{-0.33}) \cdot 10^{-10}$ and recently, $\mathcal{B}(B^+ \to K^+ \nu \bar \nu)=(2.3 \pm 0.7) \cdot 10^{-5}$ \cite{ParticleDataGroup:2024cfk}. 
In charm, no such FCNC-mode to invisibles has been seen. Only few upper limits on branching ratios have been reported  by BESIII \cite{BESIII:2024rkp,BESIII:2021slf,BESIII:2022vrr} and  Belle \cite{Belle:2016qek} at the level of  $10^{-5}-10^{-4}$.
On the other hand, new physics (NP)-induced branching ratios in charm can be large, especially 
 if the quarks are $SU(2)_L$-singlets, or   NP is light,
as in both cases the strong down-type quark constraints can be  evaded.
 In addition, the invisible searches are mostly subjected to kinematic cuts, and therefore have some degree of model-dependence. That means that rates in excess of even
 $10^{-5}-10^{-4}$ are apriori not excluded, and the decays are suitable for  experimental searches at present experiments Belle II and BESIII, and future facilities 
 \cite{Achasov:2023gey,Ai:2024nmn,FCC:2025lpp}.
 A  survey of the reach of NP models, taking into account most recent theoretical and experimental developments, is therefore well motivated.

We  consider  SM extensions with heavy and light particles that lead to signatures with missing energy in charm.
Dedicated studies in Standard Model Effective Field Theory (SMEFT)  are available \cite{Bause:2020xzj},  
and harvest the rare decay - high $p_T$ frontier and synergies arising from $SU(2)_L$, linking left-handed charged leptons to the neutrinos.
We also consider LNV-contributions, as well as those with light, sterile neutrinos. We furthermore analyze rare charm decays to axion-like particles (ALPs) and dark photons.
For previous works,  see   \cite{Beltran:2023nli,Li:2023sjf,Geng:2022kmf,Carmona:2021seb,Eguren:2024oov,Faisel:2020php,Gabrielli:2016cut,MartinCamalich:2025srw,Su:2020yze,Bauer:2021mvw,Berezhiani:1989fs,Berezhiani:1990wn,Berezhiani:1990jj,Badin:2010uh}.
We aim at a systematic, comparative study of two-, three, and four-body decays of charmed mesons and baryons and  opportunities from global analysis.
 As searches typically involve kinematic cuts, we perform recasts which allow us to derive novel limits from existing data.

The plan of the paper is as follows:
The effective field theory (EFT) and light new physics models are introduced in Sec.~\ref{sec:models}.
We discuss the decays of charm hadrons  to invisibles and their observables in Sec.~\ref{sec:obs}.
We perform a recast of existing searches for charm to invisibles in Sec.~\ref{sec:recast}.
In Sec.~\ref{sec:EC} we work out experimental constraints on the parameters, masses and couplings of the models. 
The main results of this work, predictions for achievable branching ratios in the EFT and light BSM models as well as correlations are presented in Sec.~\ref{sec:UpperLimitsBRs}.
In Sec.~\ref{sec:con} we conclude.
Auxiliary information on form factors is given in the appendix.

\section{Models}
\label{sec:models}

We present  the NP models  that lead to signatures with missing energy in charm.
In Sec.~\ref{sec:SMEFT} we briefly review   the SMEFT framework, including also  LNV operators.
The SMEFT amended by right-handed (RH) neutrinos as new light degrees of freedom, the  $\nu$SMEFT, is discussed in Sec.~\ref{sec:SMNEFT}.
In Sec.~\ref{sec:WET} we consider the weak effective field theory (WET), suitable to  compute  low-energy decay observables. We also give the matching of the SMEFT and $\nu$SMEFT operators 
onto WET in Sec.~\ref{sec:match}.
Axion-like particles (ALPs) are studied in  Sec.~\ref{sec:ALPs} and light $Z^\prime$'s decaying to the dark sector in Sec.~\ref{sec:Zprime}.
Both lead to  decay structure significantly different  from the  left-handed (LH) neutrinos of the SMEFT.
Observables connected to these models are discussed in Sec.~\ref{sec:obs}.

\subsection{SMEFT}
\label{sec:SMEFT}

The SMEFT accounts for  heavy NP  consistent with Lorentz and $SU(3)_C\times SU(2)_L \times U(1)_Y$  gauge symmetry
and linear breaking of the electroweak symmetry.
The Lagrangian reads,
\begin{equation}
  \mathcal{L}_{\mathrm{SMEFT}} = \mathcal{L}_{\mathrm{SM}} + \sum_{d}^{\infty} \sum_i \frac{\mathcal{C}^{(d)}_i}{\Lambda^{d-4}} \mathcal{O}^{(d)}_i\:,
\end{equation}
where $\Lambda$ is the scale of NP that is assumed to be sufficiently separated from the weak scale given by the vacuum expectation value  of the Higgs,
 $v = (\sqrt{2} G_F)^{-1/2}\approx 246\GeV$.
The Wilson coefficients $\mathcal{C}^{(d)}_i$ of the $d$-dimensional operators
$\mathcal{O}^{(d)}_i$ parametrize the NP contribution and could be inferred from a UV theory, or experimentally extracted from fits to data.

At leading order with dimension six in the Warsaw basis \cite{Grzadkowski:2010es} the operators
\begin{align}
  \mathcal{Q}_{\ell q}^{(1)} &= \overline{Q} \gamma_\mu Q\,\overline{L}\gamma^\mu L\,,\quad \mathcal{Q}^{(3)}_{\ell q} = \overline{Q} \gamma_\mu \tau^a Q\, \overline{L} \gamma^\mu \tau^a L\:,\\
  \mathcal{Q}_{\ell u} &= \overline{U} \gamma_\mu U\,\overline{L}\gamma^\mu L\,,\quad \mathcal{Q}_{\ell d} = \overline{D} \gamma_\mu D\, \overline{L} \gamma^\mu  L
\end{align}
contribute to $c\to u \nu\overline{\nu}$ transitions, where $Q$($L$) denote the left-handed quark(lepton) $SU(2)_L$-doublet and $U$($D$) the right-handed up-type(down-type) quark singlet respectively.
Pauli-matrices are denoted as $\tau^a$ and for convenience we suppressed both quark and lepton flavor indices.

Also LNV-transitions can be probed by missing energy searches.
$\Delta L=2$ requires SMEFT operators of uneven dimension with the leading contribution to rare charm decays arising at dimension seven.
A single dimension seven operator in the basis of \cite{Lehman:2014jma,Hamoudou:2022tdn,Bause:2020xzj,Liao:2016hru},
\begin{equation}\label{eq:LNVoperator}
  \mathcal{O}_{\ell^2quH}^{prst} =  \epsilon^{\alpha\beta} \,\left(L^T_{p\alpha}\, C \,L_{r\sigma}\right)\,\left(\overline{Q}_s^{\sigma} \,U_t\right) \,H_\beta 
\end{equation}
is present to give a contribution to $c\to u \nu \nu$. Here $C$ denotes charge conjugation and the $H$ is the Higgs $SU(2)_L$-doublet.
We use greek letters for $SU(2)_L$-indices and latin letters for flavor indices here.
This is in slight contrast to $d_s\to d_t \nu\nu$ transitions, where two operators at dimension seven are present, namely
\begin{equation}
  \begin{aligned}
  \mathcal{O}_{\ell^2dqH}^{(1),prst} &=  \epsilon^{\alpha\beta}\epsilon^{\sigma\rho} \,\left(L^T_{p\alpha}\, C \,L_{r\sigma}\right)\,\left(\bar{D}_s \,Q_{t,\rho}\right) \,H_\beta \:,\\
  \mathcal{O}_{\ell^2dqH}^{(2),prst} &=  \epsilon^{\alpha\beta}\epsilon^{\sigma\rho} \,\left(L^T_{p\alpha}\, C \sigma_{\mu\nu}\,L_{r\sigma}\right)\,\left(\bar{D}_s \,\sigma^{\mu\nu} \,Q_{t,\rho}\right) \,H_\beta \:,
\end{aligned}\label{eq:LNVoperatorDOWN}
\end{equation}
with $\sigma^{\mu\nu}=\frac{i}{2}\left[\gamma^\mu,\gamma^\nu\right]$. A similar operator to $\mathcal{O}_{\ell^2dqH}^{(2)}$ with the field content $\{L,L,\bar{Q},U,H\}$, which is allowed by Lorentz and gauge symmetry,
vanishes because of different chiralities and the identity
\begin{equation}
  \bar{\psi}_1\,\sigma^{\mu\nu}P_{L(R)} \,\psi_2 \,\bar{\psi}_3 \,\sigma_{\mu\nu}P_{R(L)} \,\psi_4= 0 \:.
\end{equation}
Here $P_{L(R)}=\frac{1}{2}\left(1\mp \gamma_5\right)$ are the left- and right-handed projection operators.

\subsection{$\nu$SMEFT}
\label{sec:SMNEFT}
The SMEFT is amended by  right-handed sterile neutrinos $N$, see Ref.~\cite{Liao:2016qyd,Li:2020lba,Felkl:2021uxi}.
The lowest-dimensional  operators  that contribute to the decays in our work  arise  at dimension six, and read
\begin{equation} \label{eq:withN}
  \begin{aligned}
    \mathcal{O}_{QuNL}^{prst} &= \left(\bar{N}_p L_r\right)\left(\bar{Q}_s U_t\right) \:,\\
    \mathcal{O}_{uN}^{prst} &= \left(\bar{N}_p \gamma_\mu N_r\right)\left(\bar{U}_s\gamma^\mu U_t\right) \:,\\
    \mathcal{O}_{QN}^{prst} &= \left(\bar{N}_p \gamma_\mu N_r\right)\left(\bar{Q}_s\gamma^\mu Q_t\right) \:,
  \end{aligned}
\end{equation}
where $p$, $r$ are lepton-flavor indices and $s$,$t$ are quark-flavor indices.

\subsection{WET with left- and right-handed neutrinos}
\label{sec:WET}
We incorporate additional light right-handed (RH) neutrinos together with the left-handed (LH) ones of the SM in the 
WET-Lagrangian \cite{Bause:2020xzj}
\begin{equation}\label{eq:WET}
    \mathcal{L}^{\text{WET}}_{\nu_i \bar{\nu}_j} =  \frac{4 G_F}{\sqrt{2}} \frac{\alpha_e}{4\pi} \sum_k \mathcal{C}_k^{ij} \cdot \mathcal{Q}_k^{ij} + \text{h.c.}\:,
\end{equation}
with $\alpha_e$ the electromagnetic fine-structure constant and $G_F$ Fermi's constant.
The indices $i,j$ denote the neutrino flavors (mass eigenstates) and neutrinos are assumed to be Dirac fermions.
LH neutrinos allow for the four-fermion operators
\begin{equation}
  \mathcal{Q}_{L(R), L}^{ij} = (\bar{u}_{L(R)} \gamma_\mu c_{L(R)})(\bar{\nu}_{jL} \gamma^\mu \nu_{iL}) \:. \\
\end{equation}
While a SM contribution to $\mathcal{C}_{LL}^{ii}$ is induced by $Z$ penguin and box diagrams via loop-level, it is strongly
GIM and CKM suppressed in charm  and  entirely negligible for the purpose of this work.

For new light degrees of freedom such as  light RH neutrinos or to describe  LNV-contributions  we  extend the operators basis by additional four-fermion ones
\begin{equation}
  \begin{aligned}
  \mathcal{Q}_{L(R),R}^{ij} &= (\bar{u}_{L(R)} \gamma_\mu c_{L(R)})(\bar{\nu}_{jR} \gamma^\mu \nu_{iR}) \:, \\
  \mathcal{Q}_{S}^{(\prime),ij} &= (\bar{u}_{L(R)}  c_{R(L)})(\bar{\nu}_{j}  \nu_{i}) \:, \\
  \mathcal{Q}_{P}^{(\prime),ij} &= (\bar{u}_{L(R)}  c_{R(L)})(\bar{\nu}_{j} \gamma_5 \nu_{i}) \:,\\
  \mathcal{Q}_{T(T_5)}^{ij} &= (\bar{u} \sigma_{\mu\nu} c)(\bar{\nu}_{j} \sigma^{\mu\nu} (\gamma_5) \nu_{i})  \:.
\end{aligned}
\end{equation}
In the observables the neutrino flavors are not measured and it is therefore required to sum
the flavor indices incoherently in all observables.
For the branching fraction this corresponds to
\begin{equation}
  \mathcal{B}(c\to u\nu \bar \nu) = \sum_{ij} \mathcal{B}(c\to u \nu_j \bar{\nu}_i) \:.
\end{equation}
For convenience we define here the combinations of Wilson coefficients
that enter various observables connected to the $c\to u$ transitions and appear in Sec.~\ref{sec:obs}
\begin{equation}
  \begin{aligned}
	x_{SP\pm} &= \sum_{ij} \left| \mathcal{C}_{S}^{ij} \pm \mathcal{C}_{S}^{\prime, ij}  \right|^2 + \left| \mathcal{C}_{P}^{ij} \pm \mathcal{C}_{P}^{\prime, ij}  \right|^2 \, ,
    \\
	x_{LR\pm} &= \sum_{ij} \left| \mathcal{C}_{LL}^{ij} \pm \mathcal{C}_{RL}^{ij}  \right|^2  + \left| \mathcal{C}_{RR}^{ij} \pm \mathcal{C}_{LR}^{ij}  \right|^2 \, ,
     \\
	x_{T} &= \sum_{ij} \left| \mathcal{C}_{T}^{ij}  \right|^2 + \left| \mathcal{C}_{T_5}^{ij}  \right|^2 \, .
\end{aligned}\label{eq:variables}
\end{equation}

\subsection{Matching onto WET \label{sec:match}}

In the SMEFT at dimension six only the operators $\mathcal{Q}_{LL}$ and $\mathcal{Q}_{RL}$ are induced~\cite{DiCanto:2025fpk}
\begin{equation}
  \mathcal{C}_{LL} = \frac{\sqrt{2}\pi}{\alpha_e G_F \Lambda^2} \left(\mathcal{C}_{\ell q}^{(1)}+\mathcal{C}_{\ell q}^{(3)}\right) \:,  \quad
  \mathcal{C}_{RL} = \frac{\sqrt{2}\pi}{\alpha_e G_F \Lambda^2} \mathcal{C}_{\ell u} \:.
\end{equation}
At dimension seven also scalar and pseudoscalar operators $\mathcal{Q}_{S,P}$ and $\mathcal{Q}_{S,P}^\prime$ are induced.
The matching can be read-off  by separating the operator~\eqref{eq:LNVoperator} into its $SU(2)_L$ components
\begin{equation}
  \begin{aligned}
  \frac{1}{\Lambda_{\text{LNV}}^3} & \mathcal{C}_{\ell^2quH}^{prst} \mathcal{O}_{\ell^2quH}^{prst} \\
  &\rightarrow \frac{v_T}{\sqrt{2}\Lambda_{\text{LNV}}^3} \mathcal{C}_{\ell^2quH}^{prst}
  \left[ \begin{split}
    &\left(\nu^T_{Lp}\,C\,\nu_{Lr} \right)\left(\bar{u}_{Ls} u_{Rt} \right) \\
    &+ \left(\ell^T_{Lp}\,C\,\nu_{Lr} \right)\left(\bar{d}_{Ls} u_{Rt} \right)
  \end{split} \right] \: .
\end{aligned}
\end{equation}
Here $v_T$ is the vacuum expectation value (vev) of the Higgs which in our case equals the SM Higgs vev $v$.
The tree level matching onto the scalar and pseudoscalar operators in Eq.~\eqref{eq:WET} reads
\begin{equation}
  \begin{aligned}
  \mathcal{C}_{S(P)}^{ij} =
  \pm \sqrt{2} \frac{2\pi}{\alpha_e} \left(\frac{v}{\Lambda_{\text{LNV}}}\right)^3 \mathcal{C}_{\ell^2quH}^{ij12} \:, \\
  \mathcal{C}_{S(P)}^{\prime\,ij} =
  + \sqrt{2} \frac{2\pi}{\alpha_e} \left(\frac{v}{\Lambda_{\text{LNV}}}\right)^3 \mathcal{C}_{\ell^2quH}^{ji21\,\ast} \:, 
  \end{aligned}
\end{equation}
in agreement with~\cite{Liao:2020zyx,Hamoudou:2022tdn}.

In the $\nu$SMEFT we similarly perform a tree-level matching and obtain
\begin{equation}
  \begin{aligned}
    \mathcal{C}_{LR}^{ij} &= \frac{\sqrt{2}\pi}{\alpha_e G_F \Lambda_N^2}\mathcal{C}_{QN}^{ij12} \:,
    &\mathcal{C}_{S(P)}^{ij} &= \pm \frac{1}{2} \frac{\sqrt{2}\pi}{\alpha_e G_F \Lambda_N^2} \mathcal{C}_{QuNL}^{ij12} \:, \\
    \mathcal{C}_{RR}^{ij} &= \frac{\sqrt{2}\pi}{\alpha_e G_F \Lambda_N^2}\mathcal{C}_{uN}^{ij12} \:,
    &\mathcal{C}_{S(P)}^{\prime\,ij} &= + \frac{1}{2} \frac{\sqrt{2}\pi}{\alpha_e G_F \Lambda_N^2} \mathcal{C}_{QuNL}^{ji21\,\ast} \:.
  \end{aligned}
\end{equation}

To summarize, in  the $d=6$ SMEFT, SMEFT at $d=7$   with LNV and the $d=6$ $\nu$SMEFT
the following combinations of coefficients (\ref{eq:variables})  are induced
\begin{equation}
\begin{aligned}
x_{LR \pm} \propto \sum_{ij} \left|\mathcal{C}_{\ell q}^{(1)\,ij} + \mathcal{C}_{\ell q}^{(3)\,ij} \pm \mathcal{C}_{\ell u}^{ij}\right|^2,& \quad \text{SMEFT} \: d=6   \\
 x_{SP \pm} \propto \sum_{ij} \left|\mathcal{C}_{\ell^2quH}^{ij12}\pm \mathcal{C}_{\ell^2 quH}^{ji21\ast} \right|^2, & \quad  \parbox[t]{2.2cm}{$\text{SMEFT} \: d=7$\\(\text{LNV})}\\
\left.\begin{split}
x_{LR \pm}&\propto \sum_{ij} \left|\mathcal{C}_{QN}^{ij12}\pm \mathcal{C}_{uN}^{ij12} \right|^2,\\
x_{SP \pm} &\propto \sum_{ij}\left|\mathcal{C}_{QuNL}^{ij12}\pm \mathcal{C}_{QuNL}^{ji21\ast} \right|^2
\end{split} \right\}. & \quad \text{$\nu$SMEFT} \: d=6 \\
\end{aligned}
\end{equation}
There is no tensor generated, $x_T=0$ in all cases. If one turns on just a single ($\nu$)SMEFT operator,  relations arise
\begin{equation}\label{eq:variables_SMEFT}
x_{LR} \equiv x_{LR+}=x_{LR-} \ \, , \quad x_{SP}\equiv x_{SP+}=x_{SP-} \:.
\end{equation}
Since dineutrino modes are SM null tests, they are very sensitive to the NP scale. For $d=6$ branching ratios are $\propto 1/\Lambda^4$ and
$\propto 1/\Lambda^6_\text{LNV}$ for $d=7$.

\subsection{ALPs}
\label{sec:ALPs}

Axion-like particles (ALPs) are pseudo Nambu-Goldstone bosons originating from a spontaneously broken global $U(1)$ symmetry.
For the "classical" QCD axion, which arises from the Peccei-Quinn (PQ) symmetry to solve the CP-problem of QCD, the coupling to photons
and its mass are related. ALPs provide a more general framework  with additional couplings, arising in  super string theory~\cite{Witten:1984dg}, supersymmetry with the
R-axion in \cite{Bellazzini:2017neg} and in composite Higgs-models in \cite{Ferretti:2013kya}.
ALPs can also serve as a candidate for dark matter \cite{Arvanitaki:2019rax}, and arise in flavor symmetry breaking \cite{Berezhiani:1989fs}.
We work in the effective field theory framework for ALPs described in  \cite{Bauer:2020jbp,Bauer:2021mvw} with a general effective Lagrangian up to dimension five at
the UV scale $\Lambda = 4\pi f$.
The effective ALP Lagrangian coincides with the one of the  QCD axion for $m_{a}=0$.

We evaluate observables below the electro-weak scale and use the part of the effective Lagrangian,
\begin{equation}
  \begin{split}
    \mathcal{L}_{\text{ALP}}^{c\to u} = \frac{\partial^\mu a}{2f}  \Bigl( &k^V_{12}\,\bar{u} \,\gamma_\mu \,c
     + k^A_{12}\,\bar{u}\,\gamma_\mu\gamma_5\,c  \Bigr) + \text{h.c.}\:,
  \end{split}
\end{equation}
that describes the flavor off-diagonal interaction of an ALP with quarks.
Here there are two coupling constants $k^{V(A)}_{12}$, which are vector and axial-vector couplings respectively.
The flavor-changing couplings in the up-sector remain scale invariant neglecting small Yukawas (i.e. $y_b$ and smaller), see \cite{Bauer:2020jbp}.

Various decay channels of ALPs are plausible depending on the choice of couplings and their existing experimental
constraints. 
To obtain contributions to missing energy we follow the approach in Ref.~\cite{Bauer:2020jbp,Bauer:2021mvw} and
consider only  ALPs  that decay outside the detector.
The pseudo-scalar nature and the momentum dependent coupling  are distinguishing features
of this model that set it apart from the other models we consider in this work.

\subsection{\texorpdfstring{Light $Z^\prime$}{Light Z'}}
\label{sec:Zprime}
We extend the SM by a light neutral $Z^\prime$ vector boson stemming from an additional $U(1)^\prime$ gauge group.
The interactions between the light $Z^\prime$ and SM particles can be described through an effective field theory, see Ref.~\cite{Fabbrichesi:2020wbt,Jaeckel:2010ni,Eguren:2024oov}.
Without loss of generality, we assume the $Z^\prime$ to be diagonalized to its mass-eigenstate basis and neglect kinematic mixing terms $\propto A^\mu Z^{\prime}_\mu$.
For the diagonalization procedure for both massive and massless $Z^\prime$'s see Ref.~\cite{Fabbrichesi:2020wbt}.
At energies below the electroweak scale it is sufficient to consider the lowest-dimension flavor-changing operators
\begin{equation}
  \label{eq:ZprimeMassless}
  \mathcal{L}^{\text{eff}}_{Z^\prime} \supset \frac{1}{\Lambda_{\text{eff}}}
  \bar{u}\left( \mathcal{C}_D^{Z^\prime} + \gamma_5 \mathcal{C}_{D5}^{Z^\prime}\right) \sigma^{\mu\nu} \, c\, Z^\prime_{\mu\nu} + \text{h.c.} \:,
\end{equation}
and for $m_{Z^\prime}\neq 0$ in addition
\begin{equation}
  \label{eq:ZprimeMassive}
  \begin{aligned}
  \mathcal{L}^{\text{eff}}_{Z^\prime} \supset & \,\mathcal{C}_L^{Z^\prime} \bar{u}_L \gamma^\mu c_L Z^\prime_\mu
    + \mathcal{C}_R^{Z^\prime} \bar{u}_R \gamma^\mu c_R Z^\prime_\mu + \text{h.c.}\\
    &+\mathcal{C}_{V}^{Z^\prime \chi} \bar{\chi} \gamma^\mu \chi Z_\mu^\prime + \mathcal{C}_{A}^{Z^\prime \chi} \bar\chi \gamma^\mu \gamma_5 \chi Z^\prime_\mu \:,
  \end{aligned}
\end{equation}
where $\chi$ denotes dark fermions charged under $U(1)^\prime$ with Dirac mass $m_\chi$. Due to the additional
operators \eqref{eq:ZprimeMassive} the  $m_{Z^\prime}\to 0$ limit is not straightforward. This could be circumvented by
 scaling $\mathcal{C}_{L/R}^{Z^\prime}(\mathcal{C}_{V/A}^{Z^\prime \chi}) \to \frac{m_{Z^\prime}}{\Lambda_{\text{eff}}}\mathcal{C}_{L/R}^{Z^\prime}(\mathcal{C}_{V/A}^{Z^\prime \chi})$
that guarantees the correct  $m_{Z^\prime}\to 0$ limit~\cite{Eguren:2024oov}. However, such a scaling is not unique and we leave Eq.~\eqref{eq:ZprimeMassive} as is.

For convenience we employ the notation
\begin{equation}
\begin{aligned}
  x_{LR\pm}^{Z^\prime} &= \left|\mathcal{C}_{L}^{Z^\prime}\pm \mathcal{C}_{R}^{Z^\prime}\right|^2  \, , \\
  x_{D(5)}^{Z^\prime} &= \left|\mathcal{C}_{D(5)}^{Z^\prime}\right|^2 \:/\: \Lambda_{\text{eff}}^2 \, ,  \\
  x_{\mathrm{Re}(\mathrm{Im})LRD}^{Z^\prime} &= \mathrm{Re}(\mathrm{Im})\left\{ \frac{\left(\mathcal{C}_L^{Z^\prime} + \mathcal{C}_R^{Z^\prime}\right)\mathcal{C}_{D}^{Z^\prime\:\ast}}{\Lambda_{\text{eff}}}\right\}  \, ,  \\
  x_{\mathrm{Re}(\mathrm{Im})LRD5}^{Z^\prime} &= \mathrm{Re}(\mathrm{Im})\left\{ \frac{\left(\mathcal{C}_L^{Z^\prime} - \mathcal{C}_R^{Z^\prime}\right)\mathcal{C}_{D5}^{Z^\prime\:\ast}}{\Lambda_{\text{eff}}}\right\} \, ,
\end{aligned}\label{eq:variables_Zp}
\end{equation}
for the combinations of Wilson coefficients that appear in branching fractions.
The last two contributions  in (\ref{eq:variables_Zp}) originate from interference terms of the Lagrangians
(\ref{eq:ZprimeMassless}) and (\ref{eq:ZprimeMassive}), which for simplicity we  do not consider for phenomenology in this work.

The  $Z^\prime$'s contribute to missing energy signatures by either predominantly
decaying to invisibles or by possessing a sufficiently large  lifetime to decay outside the detector.
For massless or stable $Z^\prime$'s the missing energy signature is
directly generated via the two-body decay topology $c\to u Z^\prime$, where the $Z^\prime$ is on-shell.

For light masses with $m_{Z^\prime} \simeq 1\,\mathrm{GeV}$ and a dominant decay through invisibles,
the $Z^\prime$'s are produced off-shell.
Specifically, we consider for simplicity the decay $Z^\prime\to\chi\bar\chi$ to be the sole contribution
to the decay width $\Gamma_{Z^\prime}$. In this case the decay topology is that of a three-body decay.
This assumption is plausible if either the $Z^\prime$ is too light to decay into two visible SM particles, but heavy enough to decay into light NP particles,
or if couplings to SM particles are too small.
For an off-shell $Z^\prime$ with momentum $q^\mu$ the decay width of $Z^\prime\to\chi\bar\chi$ is given as
\begin{equation}
  \label{eq:width_Zprime}
  \begin{split}
  \Gamma_{Z^\prime}(q^2) &= \frac{ \sqrt{q^2} }{12\pi} \sqrt{1-\frac{4m_\chi^2}{q^2}} \Biggl( |\mathcal{C}_V^{Z^\prime \chi}|^2 \left(1 + \frac{2m_\chi^2}{q^2}\right) \\
  &\quad + \left|\mathcal{C}_A^{Z^\prime \chi}\right|^2 \left(1 - \frac{4m_\chi^2}{q^2}\right) \Biggr) \:,
  \end{split}
\end{equation}
with on-shell width  $\Gamma_{Z^\prime} \equiv \Gamma_{Z^\prime}(m_{Z^\prime}^2)$.
Fixing $\Gamma_{Z^\prime}$ and the NP masses hence constrains the Wilson coefficients $ \mathcal{C}_{A,V}^{Z^\prime \chi} $.

For the phenomenological analysis we  employ a benchmark (BM) $Z^\prime$ model
\begin{align}\label{eq:ZpBMV}
  \begin{aligned}
  \text{BM$_V$ }Z^\prime:\quad  &(m_{Z^\prime},\Gamma_{Z^\prime},m_\chi) = (1,0.1,0.1)\,\mathrm{GeV} \\
  &\text{and}\quad \mathcal{C}_A^{Z^\prime \chi} = 0 \, ,
  \end{aligned}
\end{align}
and $\mathcal{C}_V^{Z^\prime \chi} = 1.94$ using (\ref{eq:width_Zprime}).

\section{Charm decays to invisibles}
\label{sec:obs}

In this section we work out  low energy observables, the  branching ratios of rare charm decays,
in the NP models of  Sec.~\ref{sec:models}.
We discuss light LH and RH neutrino observables in Sec.~\ref{sec:dineutrino_modes},
ALP observables in Sec.~\ref{sec:ALP_modes} and light $Z^\prime$ observables in Sec.~\ref{sec:Zprime_modes}.
Heavy NP in the context of SMEFT or  $\nu$SMEFT contribute only to  specific Wilson coefficients presented  in Sec.~\ref{sec:dineutrino_modes} as
discussed by the matching in Sec.~\ref{sec:match}.

We analyze both the total branching ratios as well as the differential branching ratios of decays into missing energy, $ E_\text{miss}$.
 The branching fraction as a differential of the missing energy $ E_\text{miss}$  is related
to the branching fraction  in  the invariant mass $q^2$ of the invisible particle(s) as
$\textrm{d}\mathcal{B}/\textrm{d} E_\text{miss} = 2 m_{h_c} \textrm{d}\mathcal{B}/\textrm{d}q^2$, where $m_{h_c}$ is the mass of the decaying charm hadron.
We consider the decays $D^0 \to \textit{invisible}$, $D^{0,+}\to \pi^{0,+} + \textit{invisible}$, $D_s^+ \to K^ + \textit{invisible}$, $D^0\to \rho^0(\omega) +\textit{invisible}$, $\Lambda_c\to p + \textit{invisible}$, $\Xi_c^+ \to \Sigma^+ + \textit{invisible}$ and $D^{+(0)} \to \pi^+ \pi^{-(0)} +\textit{invisible}$.

\subsection{Dineutrino modes}\label{sec:dineutrino_modes}

We work out contributions to the various decays with missing energy in models with LH or RH neutrinos.
We factorize the branching ratios of a charmed hadron $h_c$ into hadronic final states $F$ as
\begin{equation}\label{eq:BR_dineutrino}
  \mathcal{B}(h_c\to F\,\nu\overline{\nu}) = \sum_{k=\left\{SP\pm,LR\pm,T\right\}} A_{k}^{h_c\to F} x_k \, , 
\end{equation}
where $A_{k}^{h_c\to F}$ are decay-dependent coefficients and $x_k$ are the short-hand notations of  the combinations of Wilson coefficients defined in Eq.~\eqref{eq:variables}.
The numerical values for the coefficients $A_k$ with $k \in\{SP\pm,LR\pm,T\}$ are summarized in Tab.~\ref{tab:A_coeffs}.
Results are consistent with Ref.~\cite{Bause:2020xzj} albeit employing  different form factors for $D_{(s)}\to \pi(K)$.

For a single or two final state particle we additionally calculate the differential
branching fraction via
\begin{equation}\label{eq:dBR_dq2_dineutrino}
  \begin{aligned}
    \frac{\textrm{d}\mathcal{B}(h_c\to F\nu\bar\nu)}{\textrm{d}q^2} &= \sum_{k=\left\{SP\pm,LR\pm,T\right\}} a_{k}^{h_c\to F}(q^2) \,x_k  \:, \\
  \end{aligned}
\end{equation}
where $q^2$ is the invariant mass squared of the neutrinos or more general of the invisible final state particle(s).
Here $a_{k}^{h_c\to F}$ are $q^2$ and process dependent functions related to the coefficients in Eq.~\eqref{eq:BR_dineutrino} via
\begin{equation}
  \label{eq:Ak_int}
  A_{k}^{h_c\to F} = \int_{q^2_{\text{min}}}^{q^2_{\text{max}}} a_{k}^{h_c\to F}(q^2) \textrm{d}q^2 \:,
\end{equation}
with $q^2_{\text{max}} = (m_{h_c} - m_{F})^2$.
For decays without a charged $D_{(s)}$ meson we choose the full kinematically allowed region with the lower integration limits
$q^2_{\text{min}} = 0$, while for the charged modes we perform cuts to remove $\tau$ background following Ref.~\cite{Bause:2020xzj}.

The resonance background induced via $D^+\to\tau^+(\to \pi^+ \bar\nu)\nu$,
$D_s^+\to\tau^+(\to K^+ \bar\nu)\nu$ or $D^+\to\tau^+(\to \pi^+\pi^0 \bar{\nu})\nu$, leading to the same final states as the non-resonant contribution,
can be removed with a cut~\cite{Bause:2020xzj}
\begin{equation}
  q^2 > (m_\tau^2 - m_{F}^2)(m_D^2 - m_\tau^2)/m_\tau^2 \label{eq:tau_cut} \:,
\end{equation}
while still retaining sufficient phasespace for potential NP contributions. Here $m_F = m_{\pi}$, $m_{K}$ or $2m_\pi$ respectively\footnote{
  For $D^+\to\pi^+\pi^0\nu\overline{\nu}$ a lower cut $q^2 > (m_\tau^2 - p^2)(m_D^2 - m_\tau^2)/m_\tau^2$ is 
  sufficient to  remove resonant $\tau$-background. It however depends on  $p^2$, the invariant mass of the pion pair,
  and we therefore  use the simpler $p^2$-independent cut which removes only at most an additional $5\%$ of the branching fraction.}
 and $m_\tau$ is the tau-lepton mass.
We advise experiments searching for these decays to apply similar strategies and to not neglect these modes because of the resonances.
For the numerical calculation we use the values
\begin{equation}
  \label{eq:cuts_charged}
  \begin{aligned}
  q^2_{\text{min},D^+\to \pi^+}&= 0.34 \GeV^2 \:, \\
  q^2_{\text{min},D_{s}^+\to K^+}&= 0.66 \GeV^2\:, \\
  q^2_{\text{min},D^+\to \pi^+\pi^0}&= 0.31 \GeV^2\:,
  \end{aligned}
\end{equation}
when we give results on total branching ratios.

For two hadronic final states $F_{1}$, $F_2$ we calculate the three-differential branching fraction via
\begin{equation}\label{eq:d3BR_dineutrino}
\begin{aligned}
	&\frac{\textrm{d}\mathcal{B}\left( h_c\to F_1 F_2 \nu \bar \nu \right)}{\textrm{d}q^2\textrm{d}p^2\textrm{d}\!\cos\theta_{F_1}} \\
	&\quad\quad=   \sum_{k=\left\{LR\pm,SP\pm,T\right\}} b_{k}^{h_c\to F_1F_2}(q^2,p^2,\theta_{F_1}) \,x_k  \:,
\end{aligned}
\end{equation}
where the coefficient functions $b_{k}^{h_c\to F_1F_2}(q^2,p^2,\theta_{F_1})$ now depend additionally
on $p^2$ the invariant mass squared of the two final state hadrons and $\theta_{F_1}$ the angle between
$F_1$ and the negative direction of flight of the $h_c$ hadron in the rest-frame of the two hadron system.

The coefficients $A_k^{h_c\to F_1 F_2}$ of the integrated branching in Eq.~\eqref{eq:BR_dineutrino}
are calculated via
\begin{equation}
  \begin{aligned}
  A_k^{h_c\to F_1 F_2} =& \int_{q^2_{\text{min}}}^{q^2_{\text{max}}}\!\! \int_{(m_{F_1}+m_{F_2})^2}^{(m_{h_c}-\sqrt{q^2})^2}\!\! \int_{-1}^1 \,b_k^{h_c\to F_1 F_2}(q^2,p^2,\theta_{F_1}) \\
  &\times\mathrm{d}\!\cos\theta_{F_1}\mathrm{d}p^2\mathrm{d}q^2\:,
  \end{aligned}
\end{equation}
with the appropriate cuts of Eq.~\eqref{eq:cuts_charged} for $D^+\to \pi^+\pi^0\nu\overline{\nu}$.
In $D^{0}\to \pi^+\pi^{-} \nu \overline{\nu}$ resonance backgrounds via $\tau$ decays are kinematically forbidden.

\begin{table}[h!]
  \centering
   \resizebox{0.48\textwidth}{!}{
    \setlength{\tabcolsep}{9pt}
  \begin{tabular}{cccccc}
      \hline\hline
      $h_c \to F$ & $A_{SP-}^{h_c\to F}$ & $A_{SP+}^{h_c\to F}$ & $A_{LR-}^{h_c\to F}$ & $A_{LR+}^{h_c\to F}$ & $A_{T}^{h_c\to F}$\\
                                       & $[10^{-8}]$ & $[10^{-8}]$ & $[10^{-8}]$ & $[10^{-8}]$ & $[10^{-8}]$ \\
      \hline\hline
      $D^0\to$ & $143\pm 5$ & $0$ & $0$ & $0$ & $0$ \\
      \\
      $D^0\to\pi^0$ & $0$ & $3.6\pm0.1$  & $0$ & $0.89\pm0.06$  & $8.0\pm1.3$\\
      $D^+\to\pi^+$ & $0$ & $18.2\pm0.7$ & $0$ & $3.6\pm0.2$    & $39\pm7$\\
      $D^+_s\to K^+$ & $0$ & $4.0\pm0.2$ & $0$ & $0.82\pm0.05$  & $5.9\pm1.1$\\
      \\
      $D^0 \to \omega/\rho^0$ & $0.26\pm 0.04$ & $0$ & $0.91\pm 0.14$ & $0.059\pm 0.008$ & $79\pm 16$  \\
      \\
      $\Lambda_c\to p$ & $0.84\pm0.08$ & $1.8\pm0.2$  & $1.7\pm0.1$ & $1.02\pm0.09$ &  $4.3\pm0.5$\\
      $\Xi_c^+\to\Sigma^+$ & $1.5\pm0.2$   & $3.4\pm0.3$  & $3.6\pm0.3$ & $1.9\pm0.2$   & $8.8\pm0.9$\\
      \\
      $D^0\to \pi^+\pi^-$ & n.a. & $0$ & $0.57\pm0.03$ & $0.11\pm0.01$ & n.a. \\
      $D^+\to \pi^+\pi^0$ & n.a. & $0$ & $1.93\pm0.08$ & $0.47\pm0.06$ & n.a. \\
      \hline\hline
  \end{tabular}
  }
  \caption{Coefficients $A_k^{h_cF}$  defined in  Eq.~\eqref{eq:Ak_int} for various charmed hadrons $h_c$ and hadronic final states $F$. }
  \label{tab:A_coeffs}
\end{table}

\subsubsection{\texorpdfstring{$D^0 \to \nu\overline{\nu}$}{D0 -> nu nubar}}
\label{sec:D0_to_nunubar}

For $D^0 \to \nu\overline{\nu}$  decays the coefficients $A_k^{D^0\to\nu\overline{\nu}}$ of the branching ratio in Eq.~\eqref{eq:BR_dineutrino} 
vanish except for one~\cite{Bause:2020xzj}
\begin{equation}
\begin{aligned}
\label{eq:Dnunu}
  A^{D^0\to\nu\overline{\nu}}_{SP-} = \frac{G_F^2 \alpha_e^2 f_D^2}{64 \pi^3 m_{D^0} }
          \frac{m_{D^0}^6}{(m_u + m_c)^2} \; , 
\end{aligned}
\end{equation}
where $f_D = 209.0\pm 2.4\,\mathrm{MeV}$~\cite{FlavourLatticeAveragingGroupFLAG:2021npn} is the $D$-meson decay constant, $m_{D^0}$
the mass of the $D^0$-meson and $m_{c(u)}$ are the masses of the charm(up) quark.
The contributions from vector- and axial-vector-operators are helicity suppressed by two powers
of the neutrino mass and the interference between pseudo-scalar-operators and vector- and axial-vector-operators
is suppressed by one order.
At dimension six in the SMEFT we therefore do not expect a signal for $D^0 \to \textit{invisible}$.
Requisite scalar or pseudoscalar operators are, on the other hand, induced at $d=7$ or in the $\nu$SMEFT.

\subsubsection{\texorpdfstring{$D \to \pi \, \nu \bar \nu$}{D -> pi nu nubar} and \texorpdfstring{$D_s\to K\nu\overline{\nu}$}{Ds -> K nu nubar}}
\label{sec:D0toPi0nunubar}
\begin{figure}
  \centering
  \includegraphics[width=0.48\textwidth]{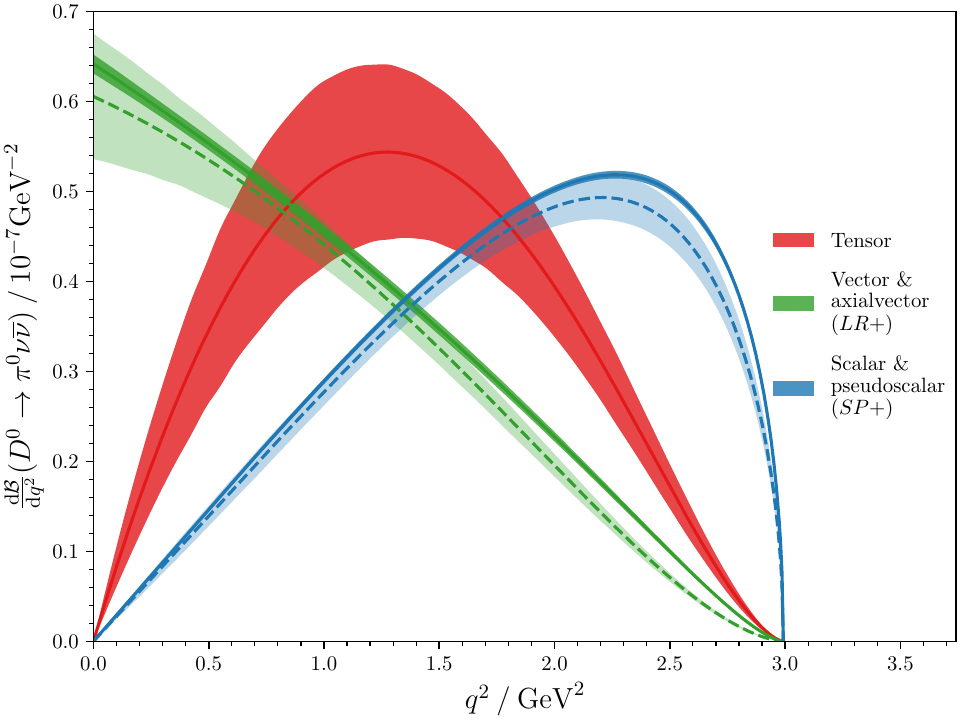}
  \caption{The $\mathrm{d}\mathcal{B}(D^0 \to \pi^0 \nu \bar \nu)/\mathrm{d}q^2$  distributions as  functions of  $q^2$.
  Solid curves $(SP+, LR+)$ are  based on the  form factors from Fermilab lattice \cite{FermilabLattice:2022gku}.
  The main source of uncertainty stems from the form factors and is illustrated  by the bands.
  Also shown  (dashed curves) are  form factors from the ETM collaboration  \cite{Lubicz:2017syv,Lubicz:2018rfs}, 
  featuring  larger uncertainties. The tensor form factor  has only been provided by ETM, and is therefore shown by the solid (red) curve.
    For each of the solid curves a single $x_k$, $k\in\{SP+,LR+,T\}$ of Eq.~\eqref{eq:variables} is turned on such that $\mathcal{B}(D^0 \to \pi^0 \nu \bar \nu ) = 10^{-7}$. 
    Identical values of  $x_k$  have been  used for the dashed and the solid curves.
    Distributions of $D^+ \to \pi^+ \nu\overline{\nu}$ are identical, see text for details.}
  \label{fig:dBR_dq2_D_to_pinunubar}
\end{figure}

For $D\to \pi\nu\overline{\nu}$ the $q^2$-dependent functions $a_{k}^{D\to\pi}$
of the differential branching fraction in Eq.~\eqref{eq:dBR_dq2_dineutrino} read \cite{Bause:2020xzj}
\begin{equation}\label{eq:D_to_p_nunubar}
  \begin{aligned}
    a_{SP+}^{D^0\to\pi^0}(q^2) &= 3 N_{D\pi}\, q^2 \frac{(m_D^2-m_\pi^2)^2}{(m_c-m_u)^2} \frac{f_0^2(q^2)}{2} \:,\\
    a_{LR+}^{D^0\to\pi^0}(q^2) &= N_{D\pi}\,\lambda_{D\pi} \frac{f_+^2(q^2)}{2} \:,\\
    a_{T}^{D^0\to\pi^0}(q^2)   &= 16 N_{D\pi}\,q^2 \frac{\lambda_{D\pi}}{(m_D+m_\pi)^2} \frac{f_T^2(q^2)}{2} \:,\\
    a_{k}^{D^+\to\pi^+}(q^2) &= 2\, a_{k}^{D^0\to\pi^0}(q^2) \:,
\end{aligned}
\end{equation}
with the normalization
\begin{equation}
    N_{D\pi} = \frac{\tau_D G_F^2\alpha_e^2 \sqrt{\lambda_{D\pi}} }{3\cdot 2^{10} m_D^3 \pi^5} \:,
\end{equation}
where $\lambda_{D\pi} = \lambda\left(m_D^2,m_\pi^2,q^2\right)$, $\lambda(a,b,c)=a^2+b^2+c^2-2(ab+ac+bc)$ is the Källén function and contributions of other Wilson coefficients vanish.
The functions $f_{+,0,T}(q^2)$ are the $D\to\pi$ form factors\footnote{
Note that we include the isospin factor $1/\sqrt{2}$ for the neutral
pion in the final state in Eq.~\eqref{eq:D_to_p_nunubar} and use the same form factors from \cite{FermilabLattice:2022gku,Lubicz:2017syv,Lubicz:2018rfs}
for $D^{0(+)}\to\pi^{0(+)}\nu\overline{\nu}$ as opposed to including it in the form factor itself.}, which are available from lattice QCD in \cite{FermilabLattice:2022gku,Lubicz:2018rfs},
see Appendix \ref{sec:FFDtopi}.
For $D_s^+ \to K^+$ we use the same expressions as for $D^+\to\pi^+$ throughout this paper with obvious kinematic replacements.
We use the same form factors for simplicity, which is supported as a good approximation by Ref.~\cite{FermilabLattice:2022gku}.
The differential distribution of $D^0\to\pi^0\nu\overline{\nu}$ decays is shown in Fig.~\ref{fig:dBR_dq2_D_to_pinunubar} by turning on a single $x_k$ coefficient for each curve and using an arbitrary normalization.
We include uncertainties from particle masses, lifetimes and form factors, with the latter being the main source of uncertainty.
One should note that only the tensor contribution always increases in the direction of the $q^2=0$ kinematic endpoint and could in this way be cleanly separated
from the other two contributions.
Distributions of $D^+ \to \pi^+ \nu\overline{\nu}$ are identical with the only differences
in the normalization caused by the lifetime and isospin factors and additional contributions of $\tau$-background in the region $q^2< 0.34\GeV^2$.

Because of discrepancies of the form factor $f_{0,+}$ in the high-$q^2$ region from different lattice calculations, see Appendix~\ref{sec:FFDtopi} and Ref.~\cite{FermilabLattice:2022gku},
we show result in solid (dashed) using most recent (ETM) form factors. The lattice form factors of Ref.~\cite{FermilabLattice:2022gku} have significantly
smaller uncertainties than those by the ETM collaboration from Ref.~\cite{Lubicz:2017syv}.
The effect of the discrepancies of the form factors on the branching ratio is partially kinematically suppressed at the endpoint and effects only a limited $q^2$ region.
At most a $\sim 6\%$ difference for the total branching fraction is observed.
For $f_T$ only a computation by the ETM collaboration is available. We emphasize to also consider $f_T$, when trying to resolve the deviations.

\subsubsection{\texorpdfstring{$D\to \rho^0(\omega) \,\nu\bar{\nu}$}{D -> rho0(omega) nu nubar}}
\label{sec:D_to_Vnunubar}
\begin{figure}
  \centering
  \includegraphics[width=0.48\textwidth]{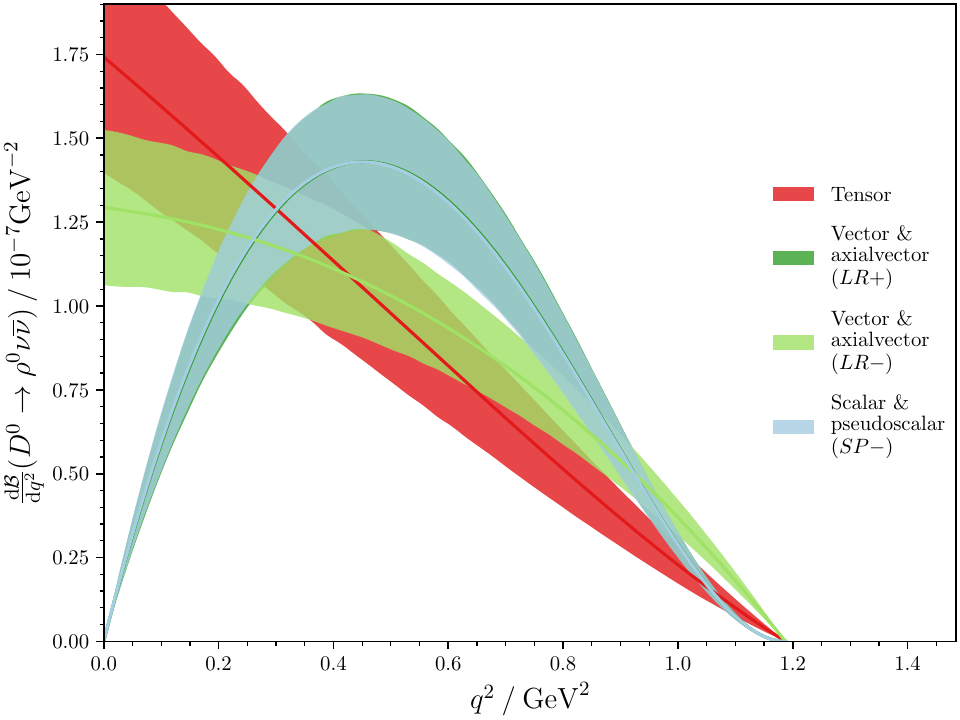}
  \caption{The $\mathrm{d}\mathcal{B}(D^0 \to \rho^0 \nu \bar \nu)/\mathrm{d}q^2$  distributions as  functions of  $q^2$, see  Fig.~\ref{fig:dBR_dq2_D_to_pinunubar}. 
  Central values are normalized to $\mathcal{B}(D^0  \to \rho^0 \nu \bar \nu ) = 10^{-7}$.
Due to the proximity of masses and  for similar form factors the  distributions for $D^0 \to \omega \nu\overline{\nu}$  are similar.
   The distributions proportional
  to $x_{LR+}$ and $x_{SP-}$ overlap, see text.}
  \label{fig:dBR_dq2_D_to_Vnunubar}
\end{figure}

For $D\to V\nu\overline{\nu}$ decays with $V=\rho^0(\omega)$ the functions $a_k^{D\to V}$ of the differential
branching fraction in Eq.~\eqref{eq:dBR_dq2_dineutrino} are given as
\begin{equation}\label{eq:DtoV_aFuncs}
  \begin{aligned}
    a_{LR+}^{D\to V}(q^2) &= 2 N_{DV} \frac{\lambda_{DV}}{(m_D+m_V)^2} V(q^2)^2 \:,\\
    a_{LR-}^{D\to V}(q^2) &= 2 N_{DV} \bigg((m_D+m_V)^2 A_1(q^2)^2 \:,\\
    &\quad+ 32\frac{m_D^2 m_V^2}{q^2} A_{12}(q^2)^2 \bigg) \:,\\
    a_{SP-}^{D\to V}(q^2) &= 3 N_{DV} \frac{\lambda_{DV}}{m_c^2} A_0(q^2)^2 \:,\\
    a_{T}^{D\to V}(q^2) &= 32 N_{DV}\bigg(\frac{T_1(q^2)^2 \lambda_{DV}+ (m_D^2-m_V^2)^2 T_2(q^2)^2}{q^2} \\
    &\quad+ 8 \frac{m_D^2 m_V^2}{(m_D+m_V)^2} T_{23}(q^2)^2\bigg) \:,\\
  \end{aligned}
\end{equation}
with the normalization
\begin{equation}
    N_{DV} = \frac{G_F^2\,\alpha_e^2\, \tau_D\, \sqrt{\lambda_{DV}}q^2}{3\cdot 2^{10} \pi^5 m_D^3} \:,
\end{equation}
where $\lambda_{DV} = \lambda(m_D^2,m_V^2,q^2)$. The scalar-/pseudoscalar contribution of $x_{SP+}$
vanishes. We find the same expressions as Ref.~\cite{Gartner:2024muk}. The form factors of the $D\to V$ transitions are defined
in Sec.~\ref{sec:FFDtoV}.
We evaluate these expression for $D^0 \to \rho^0\nu\overline{\nu}$ as well as
$D^0 \to \omega\nu\overline{\nu}$ decays, where we use as an approximation the same form factors for both. Because of
similar masses the resulting distributions are nearly identical and we give therefore no separate results. In Fig.~\ref{fig:dBR_dq2_D_to_Vnunubar} we show the differential
distribution for different combinations of Wilson coefficients. Both the scalar contributions ($SP-$) and the vector- and axial-vector
contributions ($LR+$) overlap, due to  similar kinematic dependence \eqref{eq:DtoV_aFuncs} and 
 the numerically similar $q^2$-slopes of the form factors $V$ and $A_0$.
 In absolute terms, however, the distributions differ as seen by comparing the coefficients $A_k$ in Tab.~\ref{tab:A_coeffs}.

\subsubsection{\texorpdfstring{$\Lambda_c \to p \,\nu \bar \nu$}{Lambda\textunderscore c -> p nu nubar} and \texorpdfstring{$\Xi_c^+\to \Sigma^+\nu\overline{\nu}$}{Xic -> Sigma nu nubar}}
\label{sec:Lambda_c_to_p_nunubar}
The differential branching ratio of the baryonic three-body decay $\Lambda_c\to p \nu\overline{\nu}$ is calculated
using the helicity formalism described in \cite{Gratrex:2015hna,Das:2018sms} and neglecting neutrino masses.
It reads
\begin{equation}
  \label{eq:Lambda_c_to_pnunubar}
\begin{aligned}
	\frac{\textrm{d}\mathcal{B}\left( \Lambda_c \to p\, \nu \bar \nu \right)}{\textrm{d}q^2}
	&=   \sum_{k=\left\{SP\pm,LR\pm,T\right\}} a_{k}^{\Lambda_c\to p}(q^2) \,x_k  \:,
\end{aligned}
\end{equation}
where the contributions  $a_k(q^2)$ are defined as
\begin{equation}
\begin{aligned}
	a_{SP+}^{\Lambda_c\to p}(q^2) &= 2 N_{\Lambda_c p}\, q^2 f_0^2 s_+\left( \frac{m_{\Lambda_c}-m_p}{m_c-m_u} \right)^2 \:,\\
	a_{SP-}^{\Lambda_c\to p}(q^2) &= 2 N_{\Lambda_c p}\, q^2 g_0^2 s_-\left( \frac{m_{\Lambda_c}+m_p}{m_c+m_u} \right)^2 \:,\\
  a_{LR+}^{\Lambda_c\to p}(q^2) &= \frac{2N_{\Lambda_c p}}{3}  s_- \left(f_+^2  \left(m_{\Lambda_c}+m_p \right)^2 +  2 q^2f_{\bot}^2  \right) \:,\\
	a_{LR-}^{\Lambda_c\to p}(q^2) &= \frac{2N_{\Lambda_c p}}{3}  s_+ \left( g_+^2  \left(m_{\Lambda_c}-m_p \right)^2 + 2 q^2 g_{\bot}^2 \right) \:,\\
  a_T^{\Lambda_c\to p}(q^2) &= \frac{32 N_{\Lambda_c p}}{3}  \left( h_{\bot}^2 s_- 2 \left( m_{\Lambda_c}+m_p \right)^2 + h_+^2 s_- q^2   \right. \\
    &\quad\left.+
			\tilde{h}_{\bot}^2 s_+  2 \left( m_{\Lambda_c}-m_p \right)^2 + \tilde{h}_+^2 s_+ q^2  \right) \:, \\
\end{aligned}
\end{equation}
with the normalization
\begin{equation}
  N_{\Lambda_c p} = \frac{G_F^2 \,\alpha_e^2\, \tau_{\Lambda_c}\, \sqrt{\lambda(m_{\Lambda_c}^2,q^2,m_p^2)}}{2^{11}\pi^5 m_{\Lambda_c}^3}
\end{equation}
and
\begin{equation}
  s_\pm = \left(m_{\Lambda_c}\pm m_p\right)^2 - q^2 \:.
\end{equation}
We use the form factors $f_i,g_i,h_i,\tilde{h}_i$ of the $\Lambda_c \to p$ transition in the helicity-basis introduced in Ref.~\cite{Feldmann:2011xf}.
Numerical values for the form factors are available from lattice QCD in Ref.~\cite{Meinel:2017ggx}, see Appendix \ref{sec:FFLambdaC}.
Our result agrees with calculations for $\Lambda_b \to \Lambda\ell^+\ell^-$ \cite{Das:2018sms} and $\Lambda_c\to p\ell^+\ell^-$ \cite{Golz:2021imq} in the limit of massless leptons.
For $\Xi_c^+\to\Sigma^+$ transitions we use the same form factors and the same expressions with obvious kinematic replacements.
This can be done because their respective form factors are equal in the iso-spin limit via flavor symmetries \cite{Bause:2020xzj}.
\begin{figure}
  \centering
  \includegraphics[width=0.48\textwidth]{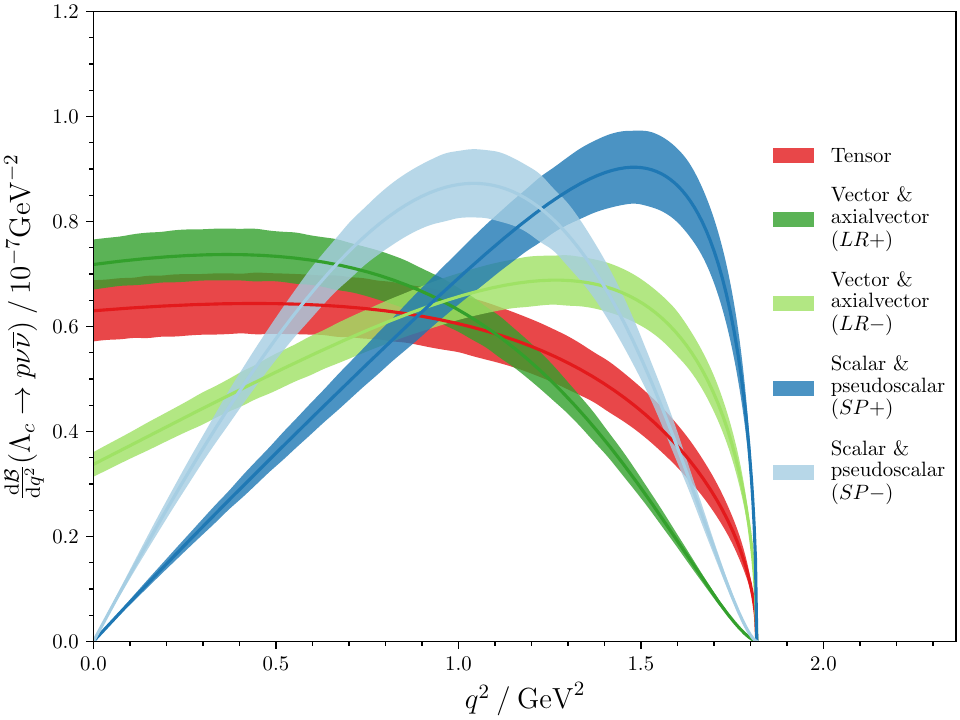}
  \caption{The $\mathrm{d}\mathcal{B}(\Lambda_c \to p \nu \bar \nu)/\mathrm{d}q^2$  distributions as functions of  $q^2$, see  Fig.~\ref{fig:dBR_dq2_D_to_pinunubar}. 
  Central values are normalized to $\mathcal{B}(\Lambda_c \to p \nu \bar \nu ) = 10^{-7}$.
  The uncertainty includes statistical and systematic errors of the form factors given in \cite{Meinel:2017ggx}.
  }
  \label{fig:DiffPlot1}
\end{figure}

In Fig.~\ref{fig:DiffPlot1} we show the differential branching fraction, where for each curve a
single $x_k$ of Eq.~\eqref{eq:variables} is turned on. All curves are normalized to $\mathcal{B}(\Lambda_c \to p \nu \bar \nu ) = 10^{-7}$.
The uncertainties of the $\Lambda_c\to p$ form factors are the main source of uncertainty for this observable and larger than for $D\to \pi$.
Fig.~\ref{fig:DiffPlot1} highlights the possibility to probe the chirality of light light neutrinos.
Scalar and pseudo-scalar NP operators, only allowed with RH neutrinos, are distinguishable by their unique $q^2$ behavior.
The curves of $x_{SP\pm}$, corresponding to scalar- and pseudo-scaler operators, vanish at the low $q^2$ endpoint, while others remain essentially finite with completely negligible corrections from neutrino masses.
Additionally, the other curves of $x_{LR\pm}$ and $x_T$ are below the scalar and pseudo-scalar contributions in the high $q^2$ region.

\subsubsection{\texorpdfstring{$D\to \pi\pi \nu\overline{\nu}$}{D -> pi pi nu nubar}}
\label{sec:D_to_pipinunubar}
The  functions $b_k$ of the three-differential branching fraction
\eqref{eq:d3BR_dineutrino} for $D^{0(+)}\to \pi^+\pi^{-(0)}\nu\overline{\nu}$ decays read
\begin{equation}
\begin{aligned}
  b_{SP-}^{D\to\pi\pi}(q^2,p^2,\theta_{\pi^+}) &= \frac{\tau_D}{2} \frac{1}{(m_c+m_u)^2}  \left|\mathcal{F}_t\right|^2 \:,\\
  b_{LR+}^{D\to\pi\pi}(q^2,p^2,\theta_{\pi^+}) &= \frac{\tau_D}{6}  \sin^2 \theta_{\pi^+} \left|\mathcal{F}_\perp\right|^2 \:,\\
	b_{LR-}^{D\to\pi\pi}(q^2,p^2,\theta_{\pi^+}) &= \frac{\tau_D}{6} \left[\left|\mathcal{F}_0\right|^2 + \sin^2 \theta_{\pi^+} \left|\mathcal{F}_\parallel\right|^2\right] \:,\\
  b_T^{D\to\pi\pi}(q^2,p^2,\theta_{\pi^+}) &=\frac{8\tau_D}{3} \bigg[\left|\mathcal{F}_0^T\right|^2 \\
  &\quad\quad+ \sin^2 \theta_{\pi^+} \left(\left|\mathcal{F}_\parallel^T\right|^2+\left|\mathcal{F}_\perp^T\right|^2\right)\bigg] \:.
\end{aligned}
\end{equation}
Here $\mathcal{F}_i(q^2,p^2,\theta_{\pi^+})$ are the transversity form factors defined in Appendix~\ref{sec:FFDtoPiPi},
 in agreement with Ref.~\cite{Bause:2020xzj}.
The scalar and tensor form factors  $\mathcal{F}_t, \mathcal{F}^T_i$  are known much less well  than the vector and axial-vector ones. We therefore refrain from  phenomenological analysis involving the former in this work.
$p^2$- and $q^2$-distributions using (axial-)vector form factors are given in \cite{DiCanto:2025fpk}.

\subsection{ALP modes}\label{sec:ALP_modes}
Decays to ALPs differ by their decay topologies compared to the scenarios described in Sec.~\ref{sec:dineutrino_modes}, \ref{sec:Zprime_modes}.
For example instead of a three-body decay for a single hadronic final state, it is a two-body decay with a delta distribution as
its differential branching fraction in $q^2$. If an experimental $q^2$ resolution is considered, smearing effects are however necessary.
The branching fraction is given as a function of the ALP mass $m_a$ via
\begin{equation}\label{eq:BR_ALP_general}
  \mathcal{B}(h_c \to F a) = \frac{\left|k^V_{12}\right|^2}{f^2} a_V^{h_c\to F}(m_a) + \frac{\left|k^A_{12}\right|^2}{f^2} a_A^{h_c\to F}(m_a)
\end{equation}
with the two possible types of ALP couplings, where $a_k^{h_c\to F}(m_a)$ parametrize the kinematic dependence on the ALP mass.

If the ALP does not decay, its contribution to signatures with invisibles is obvious. In general however ALPs can decay to various SM particles for all values of $m_a$ that we consider.
To study its contribution to missing energy modes we therefore require the ALPs to decay outside the detector.
If the decaying $h_c$ hadron is at rest, the fraction of ALPs, which escape
a detector of transverse radius  $R_{\text{max}}$, is \cite{Bauer:2021mvw}
\begin{equation}
  F_T = \int_0^{\frac{\pi}{2}} \sin\theta
  \exp\left( - \frac{m_a \Gamma_a R_\text{max}}{\left|p_{LAB}^T\right|} \right)\textrm{d}\theta
  \label{eq:FT}
\end{equation}
with the transverse momentum
\begin{equation}
  p_{LAB}^T = \sin\theta \frac{1}{2m_{h_c}} \sqrt{ \lambda(m_{h_c}^2,m_F^2,m_a^2)  } \:.
\end{equation}
The decay width $\Gamma_a$ of an ALP depends in general on further couplings of the model. For further details, see Ref.~\cite{Bauer:2021mvw}.
For the radius  we chose the outer radius of the muon identifier of the BESIII detector
$R_\text{max} = 2.8\,\text{m}$. We multiply the branching ratio by this factor when we consider $\Gamma_a \neq 0$.

\subsubsection{\texorpdfstring{$D \to \pi a$}{D -> pi a} and \texorpdfstring{$D_s\to K a$}{Ds -> K a}}
\label{sec:D_to_pi_a}
The branching fraction of $D\to \pi a$ is calculated via Eq.~\eqref{eq:BR_ALP_general} with the functions \cite{Bauer:2021mvw}
\begin{equation}
  \begin{aligned}
    a_V^{D^+\to \pi^+}(m_a) &= \frac{\tau_{D}m_D^3}{64\pi}
       \left| f_0^{D\to \pi} (m_a^2)
       \right|^2\\
       &\quad \times \left(1-\frac{m_\pi^2}{m_D^2}\right)^2 \lambda^{1/2} \left(1,\frac{m_\pi^2}{m_D^2},\frac{m_a^2}{m_D^2}\right) \:,\\
    a_V^{D^0\to \pi^0}(m_a) &= \frac{1}{2} a_V^{D^+\to \pi^+}(m_a) \:,
  \end{aligned}
\end{equation}
for both modes respectively and the axialvector contribution vanishes.
For $D_{s}^+\to K^+a$ we use the same replacements as discussed in Sec.~\ref{sec:D0toPi0nunubar}.
The branching ratio allows us to probe the vector coupling $k^V_{12}/f$,
while contributions of axial-vector couplings are unconstrained.
This is caused by the parity conservation of QCD, which causes the hadronic matrix element of
two pseudo-scalar mesons to vanish for axial currents.
\begin{figure}
  \centering
  \includegraphics[width=0.48\textwidth]{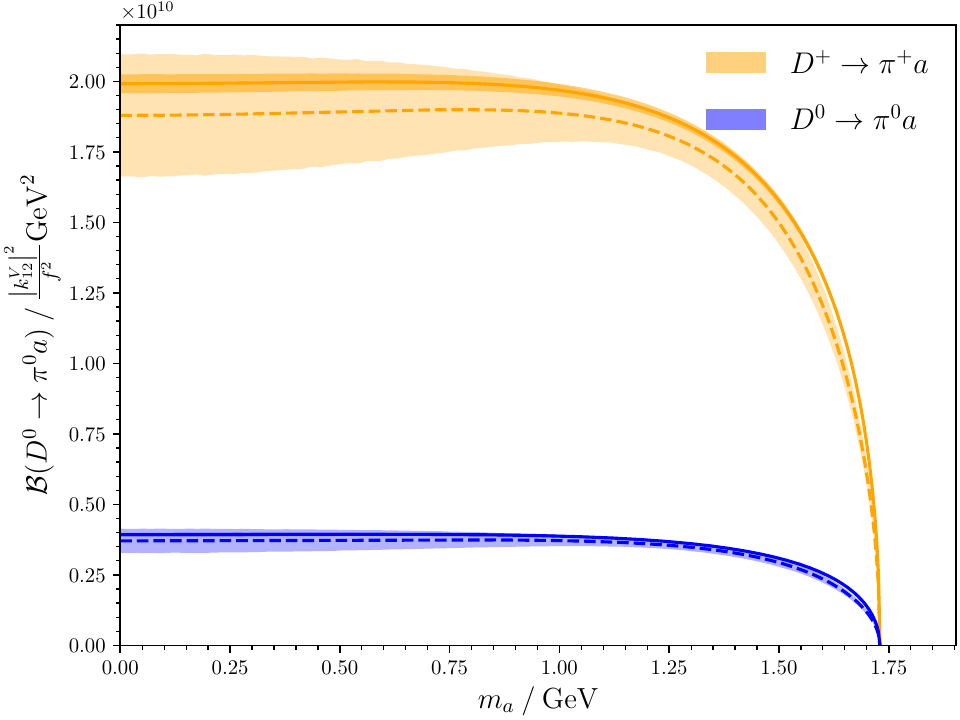}
  \caption{Branching ratio of $D^{0}\to \pi^0 a$ (lower band, blue) and $D^+\to \pi^+ a$  (upper band, orange) with  the coupling $|k^V_{12}/f|^2$ factored
  out, depending on the ALP mass $m_a$.  The difference between the charged and the neutral mode is lifetime and isospin factors. The main source of uncertainty stems from the $D\to \pi$ form factors. Solid (dashed) curves are using most recent (ETM) form factors, see  Sec.~\ref{sec:D0toPi0nunubar}.}
  \label{fig:D_ALP_Br_functions}
\end{figure}

The branching ratio of $D^{0(+)}\to \pi^{0(+)} a$
is shown in Fig.~\ref{fig:D_ALP_Br_functions} as a function of the ALP mass $m_a$.
For differential branching ratios $D\to \pi a$ differs from $D\to \pi \nu \overline{\nu}$ as its decay topology is that of a two-body decay.
The differential branching fraction would be proportional to a
Dirac delta function $\delta(q^2-m_a^2)$ and would be distinguishable from the contributions shown in Fig.~\ref{fig:DiffPlot1}.
Note that because of this the kinematic cuts for the charged modes in Eq.~\eqref{eq:cuts_charged} can not be applied for arbitrary ALP masses and only $D^0\to \pi^0+\textit{invisible}$ or baryon modes remain viable NP searches for $m_a^2 < q^2_{\text{cut}}$.

\subsubsection{\texorpdfstring{$D\to \rho^0(\omega) \,a$}{D -> rho0(omega) a}}
\label{sec:D_to_Va}
For the decay to vector mesons e.g. $D\to V a$ with $V=\rho^0,\omega$, contrary to decays to pseudo-scalar mesons,
the contribution from the vector coupling vanishes as a consequence of e.o.m.
The contribution to the branching fraction, calculated via Eq.~\eqref{eq:BR_ALP_general}, is given by
\begin{equation}
  a_{A}^{D\to V}(m_a) = \frac{\tau_D}{64\pi}m_D^3\lambda^{3/2}\left(1, \frac{m_V^2}{m_D^2},\frac{m_a^2}{m_D^2}\right) A_0^2(m_a)
\end{equation}
and probes the axial-coupling of the ALP instead. Both $D_{(s)}\to \pi(K) a$ and $D\to V$ probe therefore orthogonal directions and are complementary.

\subsubsection{\texorpdfstring{$\Lambda_c \to p a$}{Lambda\textunderscore c -> p a} and \texorpdfstring{$\Xi_c\to \Sigma a$}{Xic -> Sigma a}}
\label{sec:Lambda_c_to_p_a}
\begin{figure}
  \centering
  \includegraphics[width=0.48\textwidth]{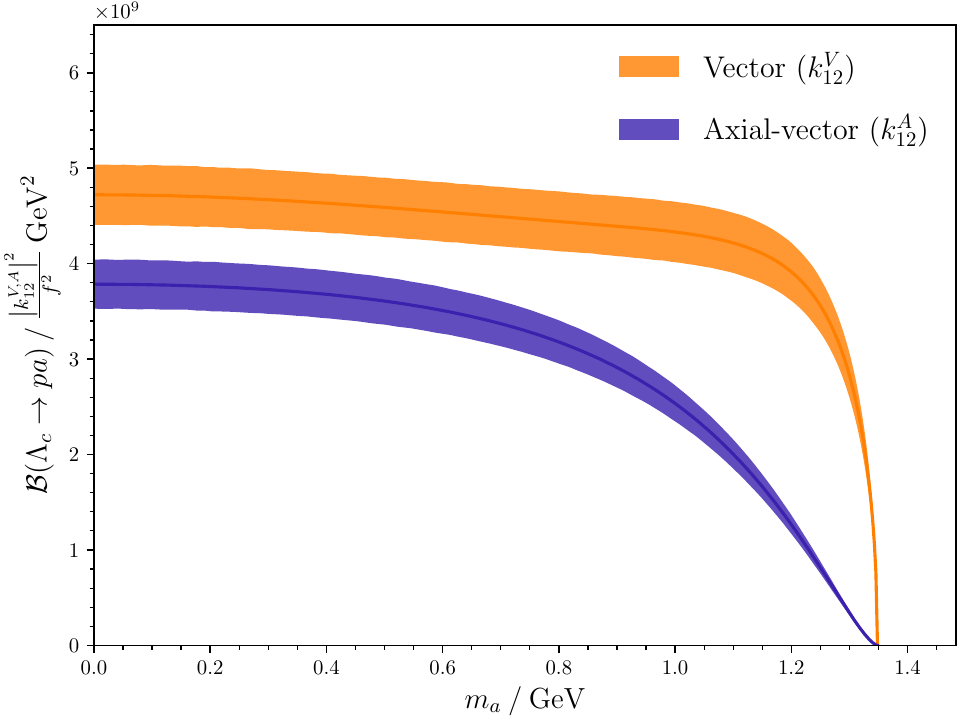}
  \caption{Branching ratio of $\Lambda_c \to pa$  against $m_a$ with  only the axial coupling switched on (lower band, blue),
  and only the vector one  turned on (upper band, orange).
  The couplings $|k^{V,A}_{12}/f|^2$ have been factored out, and the different sensitivity stems from different form factors, (\ref{eq:aVand aA}).
  The main source of uncertainty stems from the form factors, shown as bands.}
  \label{fig:normalized_BR_Lambdac_to_pia}
\end{figure}

The branching fraction of $\Lambda_c \to p a$ is calculated via Eq.~\eqref{eq:BR_ALP_general} with
\begin{equation}
  \begin{aligned} \label{eq:aVand aA}
    a_V^{\Lambda_c\to p}(m_a) &= N_{\Lambda_c p}^{\text{ALP}} (m_{\Lambda_c}-m_p)^2 s_+  f_0^2  \:,\\
  a_A^{\Lambda_c\to p}(m_a) &= N_{\Lambda_c p}^{\text{ALP}} (m_{\Lambda_c}+m_p)^2  s_-  g_0^2  \:,
  \end{aligned}
\end{equation}
and the normalization
\begin{equation}
  N_{\Lambda_c p}^{\text{ALP}}(m_a) = \frac{\tau_{\Lambda_c}  }{64 \pi m_{\Lambda_c}}
  \sqrt{\lambda\left(1,\frac{m_p^2}{m_{\Lambda_c}^2},\frac{m_{\vphantom{p}a}^2}{m_{\Lambda_c}^2} \right)}\:,
\end{equation}
where $f_0,g_0,s_\pm$ are evaluated at $q^2=m_a^2$. 
In contrast to  $D\to \pi a$  decays, $\Lambda_c\to p a$ is sensitive to  both vector and axial-vector ALP couplings.
The branching ratio as a function of the ALP mass $m_a$ is shown in Fig.~\ref{fig:normalized_BR_Lambdac_to_pia}.
Due to different form factors involved, and $f_0 > g_0$, the sensitivity to the vector coupling  is larger.
For $\Xi_c^+\to\Sigma^+$ we use the same formulas, see Sec.~\ref{sec:Lambda_c_to_p_nunubar}.

\subsubsection{\texorpdfstring{$D\to\pi\pi a$}{D->pipia}}
Similar to $D\to V a$ decays the branching ratio of $D^0\to\pi^+\pi^- a$ and $D^+\to \pi^+\pi^0 a$ is
sensitive to the axial  ALP-coupling only. The  branching ratio is obtained via  Eq.~\eqref{eq:BR_ALP_general} with
\begin{equation}
  a_A^{D\to\pi\pi} = \int_{4m_{\pi}^2}^{(m_D-m_a)^2} \mathrm{d}p^2 \int_{-1}^1 \mathrm{d}\cos\!\theta_{\pi_+}\:\frac{16 \pi^4}{\alpha_e^2 G_F^2} \frac{\sqrt{p^2}}{q^2} \left|\mathcal{F}_t\right|^2 \:.
\end{equation}
There is however insufficient knowledge
on the scalar transversity form factor  $\mathcal{F}_t$, see Appendix~\ref{sec:FFDtoPiPi}, so we refrain from using it for phenomenology in this work.

\subsection{$Z^\prime$ modes}\label{sec:Zprime_modes}

Decays that involve a $Z^\prime$ decay to invisible dark sector particles $\chi$ are, besides the couplings, also
heavily dependent on the masses and widths of the NP particles.
We consider a light $Z^\prime$ with mass $m_{Z^\prime}/2 > m_\chi$ and at most $m_{Z^\prime} \simeq \mathcal{O}(\mathrm{GeV})$.

For the on-shell production of a $Z^\prime$ in the narrow-width approximation the branching fraction of $h_c \to F Z^\prime(\to \chi \bar{\chi})$
reads
\begin{equation}
  \begin{split}
  &\mathcal{B}(h_c\to F Z^\prime(\to \chi \bar{\chi})) \\
  &\quad\quad\quad\simeq  \mathcal{B}(h_c\to F Z^\prime)(m_{Z^\prime}^2) \cdot \mathcal{B}( Z^\prime \to \bar{\chi}\chi)\:.
  \end{split}
\end{equation}
We can simplify this further assuming  that the $Z^\prime$  only decays into the dark sector, that is,  $\mathcal{B}(Z^\prime\to\chi\bar{\chi})=1$.
The result is then identical to the one for a stable $Z^\prime$.

In the case of a finite $Z^\prime$ width, where the narrow-width approximation breaks down, the branching fraction can be
approximated by \cite{Crivellin:2022obd}
\begin{equation}
  \label{eq:2body_approx}
  \begin{aligned}
  &\mathcal{B}(h_c\to F Z^\prime(\to \chi \bar{\chi})) \\
  &\quad\quad\quad= \int_{q^2_\text{min}}^{q^2_\text{max}} \frac{\mathrm{d}\mathcal{B}(h_c\to F Z^\prime(\to \chi\bar{\chi}))}{\mathrm{d}q^2}\textrm{d}q^2  \:,\\
  &\quad\quad\quad\simeq \int_{q^2_\text{min}}^{q^2_\text{max}}  \Gamma_{Z^\prime}(q^2)
   BW(q^2) \mathcal{B}(h_c \to F Z^\prime)(q^2) \textrm{d}q^2 \:,
  \end{aligned}
\end{equation}
with Breit-Wigner shape
\begin{equation}
  BW(q^2) = \frac{1}{\pi}\frac{\sqrt{q^2}}{(q^2-m_{Z^\prime}^2)^2 + \Gamma_{Z^\prime}(q^2)^2 m_{Z^\prime}^2}
\end{equation}
and $\mathcal{B}(h_c\to F Z^\prime)(q^2)$ is the off-shell two-body branching fraction with momentum transfer $q^2$.
Here,  $\Gamma_{Z^\prime}(q^2)$ is given in Eq.~\eqref{eq:width_Zprime}.
The $q^2$- integration is performed over the kinematically allowed region 
$4 m_\chi^2< q^2 \leq (m_{h_c}-m_{F})^2$ except for 
$D_{(s)}^+\to\pi^+(K^+)Z^\prime(\to \chi \overline{\chi})$, where
$q^2_{\text{min}} = \text{max}(4 m_\chi^2 ,\: q^2_{\text{min}, D_{(s)}^+\to \pi^+(K^+)} )$ to remove  the $\tau$-background \eqref{eq:cuts_charged}.

\subsubsection{\texorpdfstring{$D^0\to Z^\prime \to \chi\bar\chi$}{D0 -> Z' -> chi chibar}}
\label{sec:D0_to_Zp}
The $Z^\prime$-contribution to $D^0 \to \textit{invisible}$ reads
\begin{equation}
  \begin{aligned}
  &\mathcal{B}(D^0\to Z^\prime \to \chi \bar\chi) \\
   &\quad\quad= \frac{\tau_D
    \left|\mathcal{C}_A^\chi\right|^2 f_D^2 m_\chi^2 m_D}{8 \pi \left(
      \Gamma_{Z^\prime}^2 m_{Z^\prime}^2 + (m_{Z^\prime}^2 - m_D^2 )^2 \right)}
   \sqrt{1-\frac{4m_\chi^2}{m_D^2}}\, x_{LR-}^{Z^\prime} \:.
  \end{aligned}\label{eq:D0_to_chichibar}
\end{equation}
Similar to the EFT-contribution to dineutrinos discussed in Sec.~\ref{sec:D0_to_nunubar} a vector contribution, $\propto \mathcal{C}_V^{Z^\prime \chi}$  vanishes and
an axial-vector one, $\propto \mathcal{C}_A^{Z^\prime \chi}$  is 
suppressed by the mass  of the invisible particle, here $m_\chi$.

\subsubsection{\texorpdfstring{$D \to \pi Z^\prime(\to \chi \bar{\chi})$}{D -> pi Z'(-> chi chibar)}}
\label{sec:D_to_piZp}

The branching fraction for $D^0 \to \pi^0 Z^\prime$ with momentum-transfer $q^2$ is given as
\begin{equation}
  \label{eq:2bodyZprime}
  \begin{aligned}
    &\mathcal{B}(D^0\to \pi^0 Z^{\prime})(q^2)= \frac{\lambda^{3/2}(m_D^2,m_\pi^2,q^2)}{64 \pi m_D^3 (m_D+m_\pi)^2 q^2} \\
    &\quad\quad\times \bigg( x_{LR+}^{Z^\prime} \frac{f_+^2}{2} (m_D+m_\pi)^2 + 16 \,x_D^{Z^\prime}  \frac{f_T^2}{2} q^4 \\
    &\quad\quad+ 8 \,x_{\mathrm{Re}LRD}^{Z^\prime}  \frac{f_+ f_T}{2} q^2 (m_D + m_\pi) \bigg) \:,
  \end{aligned}
\end{equation}
which agrees with Ref.~\cite{Crivellin:2022obd,Eguren:2024oov}, for $q^2=m_{Z^\prime}^2$. We compute
the differential and total branching fraction via Eq.~\eqref{eq:2body_approx}.

\begin{figure}
  \centering
  \includegraphics[width=0.48\textwidth]{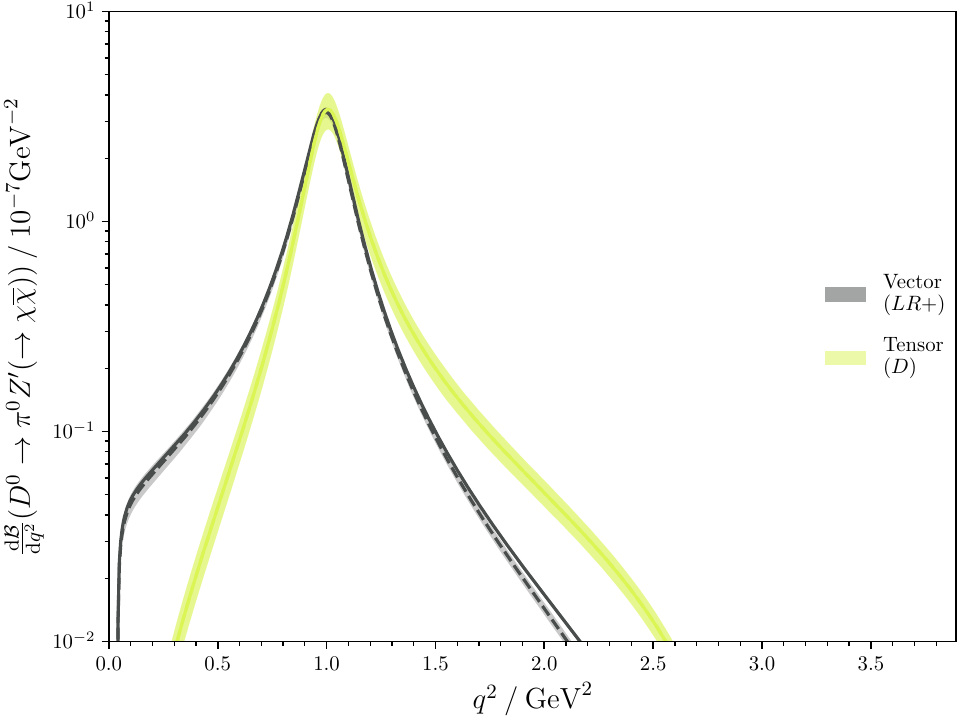}
  \caption{ Shapes of 
      $\mathrm{d}\mathcal{B}(D^0 \to \pi^0 Z^\prime(\to\chi\bar\chi))/\mathrm{d}q^2$, with  $\mathcal{B}(D^0 \to \pi^0 \nu \bar \nu ) = 10^{-7}$, for
    different couplings, obtained by switching on a single coefficient $x_k^{Z^\prime}$, $k\in\{LR+,D\}$ of Eq.~\eqref{eq:variables} for the $Z^\prime$ benchmark \eqref{eq:ZpBMV}.
  Identical for $D^+\to\pi^+ Z^\prime(\to\chi\bar\chi)$ with different normalization of the coefficients $x_k^{Z^\prime}$ caused by lifetime difference 
  and isospin factor. Solid (dashed) curves are using most recent (ETM) form factors, see Sec.~\ref{sec:D0toPi0nunubar}.
  }
  \label{fig:dBR_dq2_D_to_piZp}
\end{figure}

\begin{figure}
  \centering
  \includegraphics[width=0.48\textwidth]{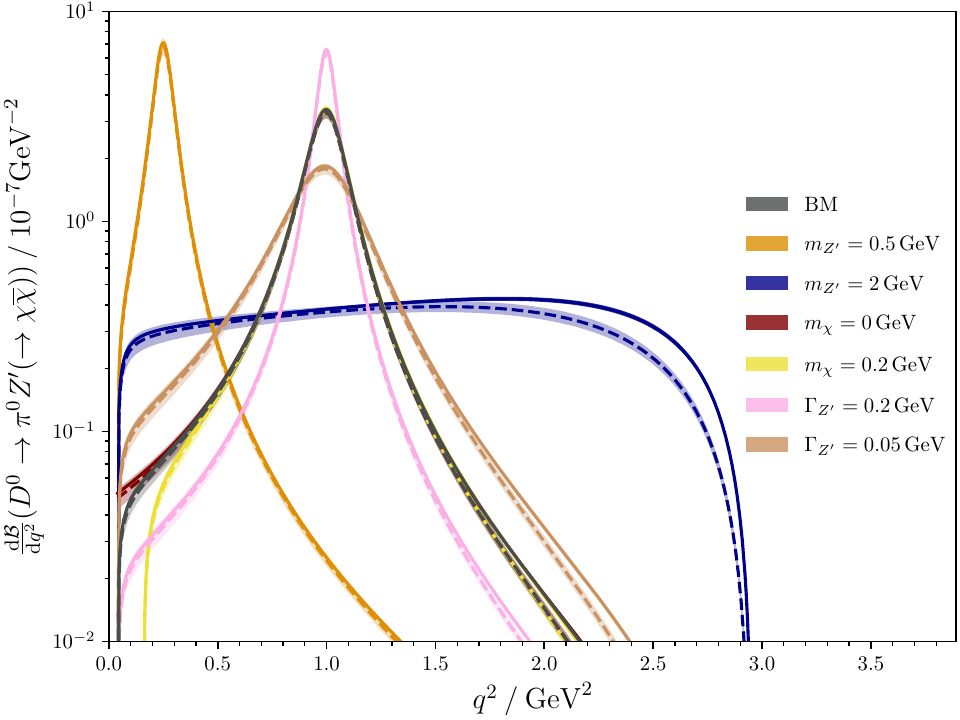}
  \caption{Shapes of 
      $\mathrm{d}\mathcal{B}(D^0 \to \pi^0 Z^\prime(\to\chi\bar\chi))/\mathrm{d}q^2$, with  $\mathcal{B}(D^0 \to \pi^0 \nu \bar \nu ) = 10^{-7}$, for
      fixed coupling  $x_{LR+}^{Z^\prime}$  but different model parameters  $m_{Z^\prime}$, $\Gamma_{Z^\prime},m_\chi$   departing from BM$_V$ of Eq.~\eqref{eq:ZpBMV}.
   }
  \label{fig:dBR_dq2_D_to_piZp_masses}
\end{figure}

In Fig.~\ref{fig:dBR_dq2_D_to_piZp} we show the differential branching ratio of $D^0 \to \pi^0 Z^\prime(\to\chi\bar\chi)$
for different couplings $x_k^{Z^\prime}$ with $k\in\{LR+,D\}$ of Eq.~\eqref{eq:variables_Zp}.
We use BM$_V$ of Eq.~\eqref{eq:ZpBMV}  and
for each of the curves a single coefficient $x_k$ is turned on such that the integrated branching ratio is  normalized to the same value.
In Fig.~\ref{fig:dBR_dq2_D_to_piZp_masses} we show the same differential branching ratio but vary instead a single model parameter  $m_{Z^\prime}$, $\Gamma_{Z^\prime}$ or $m_\chi$
and others conforming to BM$_V$. For all curves we assume only $x_{LR+}^{Z^\prime}\neq 0$ and fix its value via normalization of the branching ratio.
We learn that the shape of the differential branching ratio on the mass and width parameters of the $Z^\prime$ model is strong, while
the dependence on the coupling type is rather small.

\subsubsection{\texorpdfstring{$D \to \rho^0(\omega) Z^\prime(\to \chi \bar{\chi})$}{D -> rho0(omega) Z'(-> chi chibar)}}
\label{sec:D_to_V_Zprime}
For $D\to V Z^\prime(\to \chi\bar{\chi})$ with $V=\rho^0(\omega)$ additional couplings contribute than for
the pseudoscalar meson final states and the off-shell two-body branching fraction is given as
\begin{equation}
  \mathcal{B}(D \to V Z^\prime)(q^2) = \sum_{\substack{k=\{LR\pm,D,D5,\\\quad \mathrm{Re}LRD,\mathrm{Re}LRD5\}}} a_k^{D \to V Z^\prime}(q^2)\, x_k^{Z^\prime}
\end{equation}
with the functions
\begin{equation}
  \begin{aligned}
    a_{LR+}^{D \to V Z^\prime}(q^2) &= N_{DV}^{Z^\prime}\,  \lambda_{DV} V^2 \:,\\
    a_{LR-}^{D \to V Z^\prime}(q^2) &= N_{DV}^{Z^\prime}\,  (m_D+m_V)^2 \big(32 \frac{m_D^2 m_V^2}{q^2} A_{12}^2 \\
    &\quad\quad + (m_D+m_V)^2 A_1^2  \big) \:,\\
    a_{D}^{D \to V Z^\prime}(q^2) &= 4 N_{DV}^{Z^\prime}\, \lambda_{DV} (m_D+m_V)^2 T_1^2 \:,\\
    a_{D5}^{D \to V Z^\prime}(q^2) &= 4 N_{DV}^{Z^\prime}\, \big(8 m_D^2 m_V^2 q^2 T_{23}^2 \\
    &\quad\quad+ (m_D^2-m_V^2)^2(m_D+m_V)^2 T_2^2  \big) \:,\\
    a_{\mathrm{Re}LRD}^{D \to V Z^\prime}(q^2) &= -8 N_{DV}^{Z^\prime}\, (m_{D} + m_V) \lambda_{DV} V  T_1  \:,\\
    a_{\mathrm{Re}LRD5}^{D \to V Z^\prime}(q^2) &= 8 N_{DV}^{Z^\prime}\, (m_{D} + m_V) \big( 16 m_D^2 m_V^2 A_{12}  T_{23}  \\
    &\quad\quad+ (m_D^2-m_V^2)(m_D+m_V)^2 A_1 T_2  \big) \:,\\
  \end{aligned}
\end{equation}
and the normalization
\begin{equation}
    N_{D V}^{Z^\prime} = \frac{\lambda_{DV}^{1/2}}{64  \pi m_{D}^3 (m_D+m_V)^2} \:.
\end{equation}
For the form factors we refer to Appendix~\ref{sec:FFDtoV}.
The differential branching fraction with the subsequent decay of
the $Z^\prime$ is obtained from Eq.~\eqref{eq:2body_approx}.

\subsubsection{\texorpdfstring{$\Lambda_c \to p Z^\prime(\to \chi \bar{\chi})$}{Lambda\textunderscore c -> p Z'(-> chi chibar)} and \texorpdfstring{$\Xi_c^+\to\Sigma^+Z^\prime(\to\chi\bar{\chi})$}{Xic -> Sigma Z'(->chi chibar)}}
\label{sec:Lambda_c_to_p_Zprime}
For $\Lambda_c\to p Z^\prime$ the off-shell two-body branching fraction used in Eq.~\eqref{eq:2body_approx} for the differential
and total branching fraction reads
\begin{equation}
\begin{aligned}
  \mathcal{B}(\Lambda_c&\to p Z^\prime)(q^2) = \sum_{\substack{k=\{LR\pm,D,D5,\\\quad \mathrm{Re}LRD,\mathrm{Re}LRD5\}}} a_k^{\Lambda_c\to pZ^\prime}(q^2)\, x_k^{Z^\prime}
\end{aligned}
\end{equation}
with the functions
\begin{equation}
  \begin{aligned}
    a_{LR+}^{\Lambda_c\to pZ^\prime}(q^2) &= N_{\Lambda_c p}^{Z^\prime}\,  s_- \left(2 f_\perp^2 q^2 + f_+^2  (m_{\Lambda_c} + m_p)^2  \right) \:,\\
    a_{LR-}^{\Lambda_c\to pZ^\prime}(q^2) &= N_{\Lambda_c p}^{Z^\prime}\,  s_+ \left(2 g_\perp^2 q^2 + g_+^2  (m_{\Lambda_c} - m_p)^2  \right) \:,\\
    a_{D}^{\Lambda_c\to pZ^\prime}(q^2) &= 16 N_{\Lambda_c p}^{Z^\prime}\, s_- \left(2 h_\perp^2 (m_{\Lambda_c} + m_p)^2  + q^2 h_+^2  \right) \:,\\
    a_{D5}^{\Lambda_c\to pZ^\prime}(q^2) &= 16 N_{\Lambda_c p}^{Z^\prime}\,s_+ \left(2 \tilde{h}_\perp^2 (m_{\Lambda_c} - m_p)^2  + q^2 \tilde{h}_+^2  \right) \:,\\
    a_{\mathrm{Re}LRD}^{\Lambda_c\to pZ^\prime}(q^2) &= -8 N_{\Lambda_c p}^{Z^\prime}\, (m_{\Lambda_c} + m_p) s_- (2 f_\perp h_\perp + f_+ h_+) \:,\\
    a_{\mathrm{Re}LRD5}^{\Lambda_c\to pZ^\prime}(q^2) &= \phantom{-}8 N_{\Lambda_c p}^{Z^\prime}\, (m_{\Lambda_c} - m_p) s_+ (2 g_\perp \tilde{h}_\perp + g_+ \tilde{h}_+) \:,\\
    N_{\Lambda_c p}^{Z^\prime} &= \frac{\lambda^{1/2}(m_{\Lambda_c}^2,m_p^2,q^2)}{64 \pi m_{\Lambda_c}^3} \:,
  \end{aligned}
\end{equation}
using the form factors in Appendix~\ref{sec:FFLambdaC}.
We adopt the same formulas for $\Xi_c^+\to\Sigma^+$, see Sec.~\ref{sec:Lambda_c_to_p_nunubar}.

For massless $Z^\prime$'s, using the endpoint relation $h_\perp(0) = \tilde{h}_\perp(0)$, we recover the branching fraction from Ref.~\cite{Su:2020yze}
\begin{equation}
  \label{eq:BrZp_massless}
  \begin{aligned}
    &\mathcal{B}(\Lambda_c\to pZ^\prime) = \frac{\tau_{\Lambda_c} \left(m_{\Lambda_c}^2-m_p^2\right)^3}{2\pi m_{\Lambda_c}^3} \\
    &\quad\quad\quad\quad\quad\quad\quad\times
    h_\perp(0)^2 \frac{|\mathcal{C}_D^{Z^\prime}|^2  +|\mathcal{C}_{D5}^{Z^\prime}|^2}{\Lambda_{\text{eff}}^2}   \:.
  \end{aligned}
\end{equation}

\subsubsection{\texorpdfstring{$D\to \pi\pi Z^{\prime}(\to\chi\bar{\chi})$}{D -> pi pi Z'(->chi chibar)}}
For the decays $D^{+(0)}\to \pi^+\pi^{0(-)}Z^\prime( \to \chi\bar{\chi})$ the $q^2$-differential and total branching ratio is obtained
via Eq.~\eqref{eq:2body_approx} with the off-shell branching ratio
\begin{equation}
\begin{aligned}
  \mathcal{B}(D&\to \pi\pi Z^\prime)(q^2) = \int_{4 m_{\pi}^2}^{(m_D-\sqrt{q^2})^2} \mathrm{d}p^2 \\&\times\int_{-1}^{1} \mathrm{d}\cos\!\theta_{\pi^+}  \sum_{k} b_k^{D\to \pi\pi Z^\prime}(q^2,p^2,\theta_{\pi^+})\, x_k^{Z^\prime} \:,
\end{aligned}
\end{equation}
where $k\in\left\{LR\pm,D(5),\mathrm{Re}LRD(5),\mathrm{Im}LRD(5)\right\}$. 
The functions $b_k$ are given as
\begin{equation}
  \label{eq:a_Dtopipi_Zp_twobody}
  \begin{aligned}
    b_{LR+}^{D\to\pi\pi Z^\prime} &= \mathcal{N}^{Z^\prime}_{D\to\pi\pi} \frac{\sqrt{p^2}}{4q^2} \sin^2\theta_{\pi^+} \left|\mathcal{F}_\perp\right|^2\:, \\
    b_{LR-}^{D\to\pi\pi Z^\prime} &= \mathcal{N}^{Z^\prime}_{D\to\pi\pi} \frac{\sqrt{p^2}}{4q^2} \left(\left|\mathcal{F}_0\right|^2 + \sin^2\theta_{\pi^+} \left|\mathcal{F}_\parallel\right|^2 \right) \:,\\
    b_{D}^{D\to\pi\pi Z^\prime} &= \mathcal{N}^{Z^\prime}_{D\to\pi\pi} \sqrt{p^2}  \sin^2\theta_{\pi^+} \left|\mathcal{F}_\perp^T\right|^2  \:,\\
    b_{D5}^{D\to\pi\pi Z^\prime} &= \mathcal{N}^{Z^\prime}_{D\to\pi\pi} \sqrt{p^2}  \left(\left|\mathcal{F}_0^T\right|^2 + \sin^2\theta_{\pi^+} \left|\mathcal{F}_\parallel^T\right|^2 \right) \:,\\
    b_{\mathrm{Re}(\mathrm{Im})LRD}^{D\to\pi\pi Z^\prime} &= - \mathcal{N}^{Z^\prime}_{D\to\pi\pi}\sqrt{\frac{p^2}{q^2}} \sin^2\theta^2_{\pi^+} \mathrm{Im}(\mathrm{Re}) \left\{ \mathcal{F}_{\perp} \mathcal{F}_{\perp}^{T\,\ast} \right\} \:,\\
    b_{\mathrm{Re}(\mathrm{Im})LRD5}^{D\to\pi\pi Z^\prime} &= - \mathcal{N}^{Z^\prime}_{D\to\pi\pi}\sqrt{\frac{p^2}{q^2}} \left( \mathrm{Im}(\mathrm{Re}) \left\{ \mathcal{F}_{0} \mathcal{F}_{0}^{T\,\ast} \right\} \right.\\
    &\quad\left.+ \sin^2\theta^2_{\pi^+} \mathrm{Im}(\mathrm{Re}) \left\{ \mathcal{F}_{\parallel} \mathcal{F}_{\parallel}^{T\,\ast} \right\} \right) \:,
  \end{aligned}
\end{equation}
with the normalization factor
\begin{equation}
  \mathcal{N}^{Z^\prime}_{D\to\pi\pi} = \frac{64 \pi^4}{\alpha_e^2 G_F^2} \:.
\end{equation}
The transversity form factors $\mathcal{F}_i$ are defined in Appendix~\ref{sec:FFDtoPiPi}.

\subsection{Distinguishing models \label{sec:dist} }

\begin{figure}
  \includegraphics[width=0.49\textwidth]{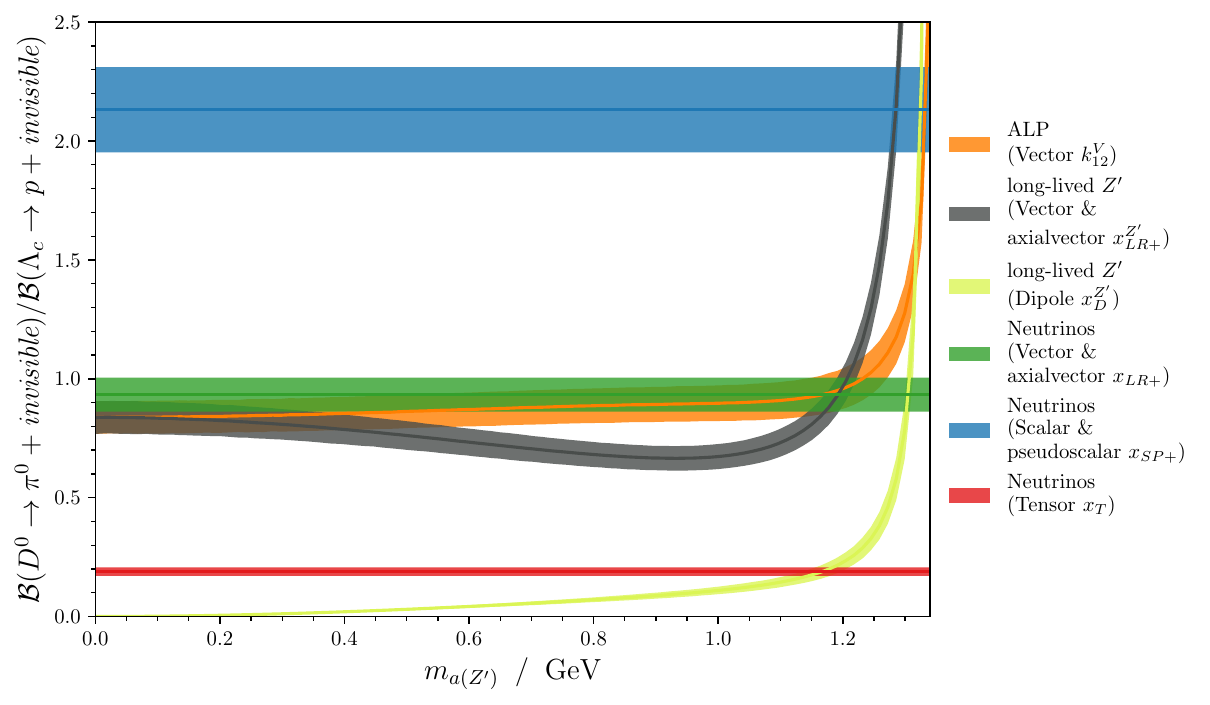}
  \caption{Ratio of the branching fractions of $D^0\to\pi^0+\textit{invisible}$ over $\Lambda_c\to p+\textit{invisible}$ for
  ALP models with $\Gamma_a = 0$ (orange) and long-lived $Z^\prime$ models with vector coupling (gray) and dipole coupling (yellow)  against the invisible mass $m_X$, $X=a$ or $X=Z^\prime$.
  Also shown by horizontal bands are EFT predictions with vector  $x_{LR+}$,  scalar $x_{SP+}$ and tensor $x_T$ structure.}
  \label{fig:correlation_ratio_ALP_Zp}
\end{figure}

We discuss  possibilities to distinguish models from decays to invisibles, for which we consider
neutrinos from contact interactions and  light new physics (ALPs, $Z^\prime$), here collectively denoted by $X$.
Depending on the width $\Gamma_X$ of the latter, we have the following cases:

$\Gamma_X =0$, that is, infinite lifetime. The decays such as $D \to \pi X$ are 2-body decays.
One can extract the mass of $X$, $m_X$, from the missing energy, $m_X^2=2 m_D E_{miss}+m_\pi^2-m_D^2$.
To distinguish different invisibles with similar mass needs more than one decay. ALPS, $Z^\prime$ have different couplings to the SM,
and due to spin come with different form factors. This induces in general different branching ratios for identical masses.
If the mass is extracted, and the branching ratio measured, the couplings can be obtained, in a given model, which provides correlations with and predictions for  other decay modes.
(Up to complications if more than one coupling per model is active. In that case, say with two couplings as in the ALP model  one would need to measure two modes and then predict a third one ).

In Fig.~\ref{fig:correlation_ratio_ALP_Zp} we show the ratio of $\mathcal{B} (D \to \pi X)$  to $\mathcal{B} (\Lambda_c \to p X)$
 in  the EFT, ALP, and  $Z^\prime$ models which induce $D \to \pi X$ decays.
Scenarios with $\mathcal{B} (D \to \pi X)=0$, e.g. $x_{LR-}\neq 0$ or the axial-vector coupling of ALPs, induce  $B (\Lambda_c \to p X)$, and can be correlated with $D \to V X$, or $D \to \pi \pi X$.
 We learn that correlations can
distinguish  the models.

 The other extreme case is that $X$ decays promptly, perhaps predominantly to dark fermions. This gives rise, for instance  in $D \to \pi (X \to \chi \chi)$
to a  3-body decay with a distribution in missing energy.
This induces four-fermion operators, with different Dirac structure, that can give different shapes shown, for instance, in Fig.~\ref{fig:dBR_dq2_D_to_pinunubar}.
Also here the models can be distinguished.

The intermediate case is when the $X$ is a resonance. Its peak gives $m_X$. For similar mass, the rates  and shapes of the distributions differ, as illustrated in 
Fig.~\ref{fig:dBR_dq2_D_to_piZp_masses}.

\section{Recast of experimental data}
\label{sec:recast}

\begin{table}[t]
  \setlength{\tabcolsep}{6pt}
  \centering
  %\captionsetup{format=plain}
  \caption{Experimental limits on rare charm decays with dineutrinos or invisible BSM particles in the final state.\\
   $^\dagger$With $\tau^+$ decaying to $\bar \nu$ plus a  light meson.}
  \label{tab:ExperimentalInputs}
  \centering
  \renewcommand{\arraystretch}{1.2}
  \begin{tabular}{
    l
    c
}
\hline
\hline
Decay & Exp. limit $@90\,\%$ C.L.  \\
\hline
$\mathcal{B}\left(D^0\to \textit{invisible}\right)$ & $<9.4\cdot 10^{-5}$~\cite{Belle:2016qek} \\
$\mathcal{B}\left(D^0\to \pi^0 \nu\bar\nu\right)$ & $<2.1\cdot 10^{-4}$~\cite{BESIII:2021slf} \\
$\mathcal{B}\left(D^0\to \omega \gamma^\prime\right)$ & $<1.1\cdot 10^{-5}$~\cite{BESIII:2024rkp} \\
$\mathcal{B}\left(D^+\to \tau^+ \nu\right)^\dagger$ & $< 1.2\cdot 10^{-3}$~\cite{CLEO:2008ffk} \\
$\mathcal{B}\left(\Lambda_c\to p \gamma^\prime\right)$ & $<8.0\cdot 10^{-5}$~\cite{BESIII:2022vrr} \\
\hline
\hline
\end{tabular}
\end{table}

Upper bounds on branching ratios of rare charm decays with invisibles in the final state are listed in Tab.~\ref{tab:ExperimentalInputs}.
We discuss how we perform a simplified reanalysis ("recast") of the experimental data under a different signal shape. This is
necessary to reinterpret the bounds for the different models we consider.

Experimental collaborations measure upper limits on the branching ratio via a bin-wise distribution of the differential branching ratio in the invariant mass of the invisible particle(s) $q^2$ or
some otherwise comparable kinematic variable. Under optimal conditions the number of bin-wise background events $b_i$ and efficiencies $\epsilon_i$, as well as the performed kinematic cuts and considered smearing effects are provided.
In this case the number of signal events $s_i$ in a $q^2$-bin $i$ is given as~\cite{Eguren:2024oov}
\begin{equation}
  s_i = \tilde{\mathcal{B}}_i(h_c\to F+\textit{invisible}) \times N_{\text{tot}} \times \epsilon_i \:,
\end{equation}
where $N_{\text{tot}}$ is the total number of $h_c$ hadrons and $\tilde{\mathcal{B}}_i$ denotes the branching ratio within bin $i$ including
other kinematic cuts and smearing. For our likelihood
\begin{equation}
  \begin{aligned}
  &\mathcal{L}(\mathcal{B},\vec{\nu} | \vec{n})= \\ &\quad\quad\prod_{i=1}^{m} \text{Pois}\bigg(n_i; s_i(\mathcal{B},\vec{\nu}) + b_i(\vec{\nu})\bigg)
  \times \prod_{i}^{k} \mathcal{N}\bigg(\nu_i;\mu_{\nu_i},\sigma_{\nu_i}\bigg)
  \end{aligned}\label{eq:likelihood}
\end{equation}
we consider a product of Poisson distributions $\text{Pois}(n;\lambda)=\lambda^n e^{-\lambda}/n!$ for the binned measurements and Gaussian
distributions $\mathcal{N}(x;\mu,\sigma)=e^{-\frac{(x-\mu)^2}{2\sigma^2}}/\sqrt{2\pi\sigma^2}$ for nuisance parameters $\vec{\nu}$ like the number of background events etc. Here $\mu_{\nu_i}(\sigma_{\nu_i})$ are the expectation value (standard deviation) of the
nuisance parameter $\nu_i$. Recast limits on the branching ratio $\mathcal{B}$ are obtained by performing a profile likelihood ratio test and
by assuming for simplicity that the test statistic is $\chi^2$ distributed.

For the measurement of $\mathcal{B}\left(D^0\to \pi^0 \nu\overline{\nu}\right)$ by Ref.~\cite{BESIII:2021slf} the signal region is  $q^2\in\left[1.1,1.9\right]\,\mathrm{GeV}^2$ and
no bin-wise efficiencies and no proper\footnote{In Ref.~\cite{BESIII:2021slf} a per-bin MC background has been provided which however does not directly agree with data
and has been adjusted by a corrections factor for the whole signal region.} bin-wise background has been provided. The used signal shape corresponds to our scenario with light LH and RH neutrinos
with only the coefficient $x_{LR+}$ non-zero.
To reinterpret this result we therefore can follow a more simplified approach, assuming no bin-dependence of the efficiencies and by translating the bound on the full branching ratio
to a bound on the branching ratio in the bin $q^2\in\left[1.1,1.9\right]\,\mathrm{GeV}^2$. We obtain
\begin{equation}
  \label{eq:recast_D0_to_pi0}
  \langle \mathcal{B} \rangle (D^0 \to \pi^0 +\textit{invisible}) \bigg|_{q^2\in\left[1.1,1.9\right]\,\mathrm{GeV}^2} \lesssim 5.8 \cdot 10^{-5}
\end{equation}
and constrain other signal shapes by evaluating them in the same $q^2$ bin.

The measurement of $D^+ \to \tau^+ \nu_\tau$~\cite{CLEO:2008ffk} can be recast to a measurement of $D^+\to \pi^+ + \textit{invisible}$ within the signal region
$q^2 < 0.05\,\mathrm{GeV}^2$~\cite{MartinCamalich:2020dfe}.
The number of measured signal events is $n=11$ and the estimated number of  background events  is  $b=13.5\pm1.0$ in the signal region.
With the total number of tagged $D$ decays $N_{\text{tot}} = 4.6\cdot 10^{5}$ and single pion detection efficiency $\epsilon=0.89$ this yields~\cite{MartinCamalich:2020dfe} 
\begin{equation}
\label{eq:cleo-tau}
\langle\mathcal{B}\rangle(D^+\to\pi^++\textit{invisible})\bigg|_{q^2<0.05\,\mathrm{GeV}^2} \lesssim 8.0\cdot 10^{-6}
\end{equation} at $90\%\: C.L.$, based on the mixed bayesian-frequentist approach given in Ref.~\cite{Barlow:2002bk}.
Here,  the  Poisson distributed part of the likelihood in Eq.~\eqref{eq:likelihood} (with the number of bins $m=1$) is maximized with an  estimator $Q$  and then 
a subsequent bayesian analysis of $Q$  is performed. The estimator $Q$ is defined as~\cite{Barlow:2002bk}
\begin{equation}
  Q= \frac{n\, N_{\text{tot}} \epsilon}{N_{\text{tot}} \epsilon \langle \mathcal{B} \rangle_{\text{data}}+b} - N_{\text{tot}} \epsilon\,,
\end{equation}
with $ \langle \mathcal{B} \rangle_{\text{data}}$ fixed by
\begin{equation}
\frac{n\, N_{\text{tot}} \epsilon}{ N_\text{tot} \epsilon \langle\mathcal{B}\rangle_{\text{data}}+E[b]} - N_\text{tot} \epsilon = 0
\end{equation}
and $E[b]=13.5$, the expectation value of $b$. The bayesian analysis is performed by generating a normal distributed
sample of background events,  $\{b\}$, and subsequently a sample of observed events, $\{n\}$
for each tested value of $\langle \mathcal{B} \rangle$ using a Poisson distribution with the expectation values $\{\mu\} = N_{\text{tot}} \epsilon \langle \mathcal{B} \rangle +\{b\}$.
The limit on the branching ratio at a $C.L.$ of $\alpha$ corresponds to the value of $ \langle \mathcal{B} \rangle$  for which the fraction $\alpha$ of the samples has positive $Q$.

Following instead  a purely frequentist approach by performing a profile likelihood ratio test with the likelihood in Eq.~\eqref{eq:likelihood}
using minimization routines from \textit{iminuit}~\cite{iminuit}, we obtain  an upper limit on the branching ratio as  $6.0\cdot 10^{-6}$ at $90\%\: C.L.$ To be conservative in the statistical
approach we decide to use the bound of Eq.~\eqref{eq:cleo-tau} in the  subsequent analysis.

The situation for  $\mathcal{B}(\Lambda_c \to p \gamma^\prime)$ and $\mathcal{B}(D^0 \to \omega \gamma^\prime)$ 
is less clear-cut, as no per-bin  efficiencies
are provided. However, the modes are useful as they probe couplings that are not accessible with $D \to \pi + \textit{invisible}$.
Experiments do provide 
bin-wise background and observed events, albeit small signal windows.  For $\Lambda_c\to p \gamma^\prime$
a signal window has been specified as $q^2 \in \left[0.0,0.1\right] \mathrm{GeV}^2$, while for $D^0 \to \omega \gamma^\prime$ no such signal region is given.
In both cases the signal shape has been smeared by a gaussian distribution, we verified however that for our application this effect is rather small $( < 1\%)$ and will not include
any smearing effect for the evaluated observables.
To obtain bounds for signal distributions that differ from experimental signal distributions we require 
the branching ratio in the region $q^2 \in \left[0.0,0.1\right] \mathrm{GeV}^2$ to fulfill the bounds in Tab.~\ref{tab:ExperimentalInputs}
for the dark photon.
We do this  for both modes although the bound for $D^0 \to \omega \gamma^\prime$ uses potentially a bigger region. This should be done because
also for $D^0 \to \omega \gamma^\prime$ the signal lies mostly within $q^2 \in \left[0.0,0.1\right] \mathrm{GeV}^2$ and the bound given in Ref.~\cite{BESIII:2024rkp}
should approximately apply in this region. If we would consider signals with higher contributions outside this window we expect higher deviations and
a more detailed bin-wise recast is necessary but beyond the scope of this work.
We conservatively use
\begin{equation} \label{eq:darkphoton}
\begin{aligned}
\langle\mathcal{B}\rangle(\Lambda_c \to p +\textit{invisible})\bigg|_{q^2<0.1\,\mathrm{GeV}^2} & \lesssim 8.0\cdot 10^{-5} \:, \\
\langle\mathcal{B}\rangle(D^0 \to \omega +\textit{invisible})\bigg|_{q^2<0.1\,\mathrm{GeV}^2} & \lesssim 1.1\cdot 10^{-5} \:.
\end{aligned}
\end{equation}

We learn that for two-body decays into an invisible particle, such as a  long-lived $Z^\prime$ or an ALP,  two mass windows  can be probed with present data,
a low mass region $m_a < 0.22$ GeV using (\ref{eq:cleo-tau}), (and $m_a < 0.3$ GeV using (\ref{eq:darkphoton}))  and a high mass window $1.05 <m_a < 1.38$ GeV using the dineutrino search (\ref{eq:recast_D0_to_pi0}).

Outside these mass windows, and more general for decay modes of all charmed hadrons,
 lifetime constraints  apply. They are obtained from  subtracting the sum of all reported  exclusive
branching ratios from unity.  Using \cite{ParticleDataGroup:2024cfk} we find for the  unaccounted branching ratios
\begin{equation}\label{eq:BRs_unaccounted}
  \begin{aligned}
    \mathcal{B}(D^0\to \text{unacc.}) &\lesssim 0.050\:, & \mathcal{B}(D^+ \to \text{unacc.}) &\lesssim 0.082\:, \\
    \mathcal{B}(D_s^+\to \text{unacc.}) &\lesssim 0.39\:, &  & \\
    \mathcal{B}(\Lambda_c\to \text{unacc.}) &\lesssim 0.25\:, &\mathcal{B}(\Xi_c^+ \to \text{unacc.}) &\lesssim 0.77\:.
  \end{aligned}
\end{equation}
which are weak, however, provide constraints beyond the searches Eqs.~(\ref{eq:recast_D0_to_pi0})-(\ref{eq:darkphoton}).
As any measurement is expected to give better bounds, we will not use (\ref{eq:BRs_unaccounted}) in the tables with 
constraints on model parameters.

\section{Constraints in EFT and models \label{sec:EC}}

We work out limits on Wilson coefficients in Sec.~\ref{sec:UpperLimits} and for couplings of ALP and light $Z^\prime$ models in Sec.~\ref{sec:alp-param} and Sec.~\ref{sec:Zp-param}, respectively.

\subsection{Upper limits on Wilson coefficients}
\label{sec:UpperLimits}
\begin{table}[t]
  \setlength{\tabcolsep}{2pt}
  \centering
  %\captionsetup{format=plain}
  \caption{Upper limits from the decays $h_c\to F+\textit{invisible}$ on the combinations of Wilson coefficients $x_k$
  of the LH \& RH neutrino model in Eqs.~(\ref{eq:variables},\ref{eq:variables_SMEFT}).
  Limits are obtained using experimental input from Tab.~\ref{tab:ExperimentalInputs}. ``-'' indicates that the coupling can not be constrained by the decay. The columns with $x_{SP}$ and $x_{LR}$ are single SMEFT coefficient limits, see Eq.~\eqref{eq:variables_SMEFT}.}
  \label{tab:xCoeff_limits}
  \centering
  \renewcommand{\arraystretch}{1.2}
  %\renewcommand{\arraystretch}{0.8}
  %\resizebox{0.49\textwidth}{!}{
  \begin{tabular}{
    l
    c
    c
    c
    c
    c
    c
    c
}
\hline
\hline
& \multicolumn{7}{c}{Light LH \& RH neutrino} \\
\cmidrule[0.5pt](l{0.75em}r{0.75em}){2-8}
   & $x_{SP-}$ & $x_{SP+}$ & $x_{SP}$ & $x_{LR-}$ & $x_{LR+}$ & $x_{LR}$ & $x_T$ \\
$h_c\to F$  & $[10^0]$ & $[10^3]$ & $[10^3]$ & $[10^4]$ & $[10^4]$ & $[10^4]$ & $[10^3]$ \\
\hline
\hline
$D^0\to $        & $66$           & -      & $0.066$& -     & -      & -      & -      \\
\\
$D^0\to \pi^0$   & -              & $4.6$  & $4.6$  & -     & $2.2$  & $2.2$  & $1.8$  \\
$D^+\to \pi^+$   & -              & $112$  & $112$  & -     & $0.52$ & $0.52$ & $18.6$ \\
\\
$D^0 \to \omega$ & $13\cdot 10^4$ & -      & $133$  & $9.3$ & $58$   & $0.92$ & $0.083$\\
\\
$\Lambda_c \to p$& $15\cdot 10^5$ & $1185$ & $659$  & $129$ & $10.9$ & $5.9$  & $2.9$  \\
\hline
\hline
Best overall & $66$ & $4.6$ & $0.066$ & $9.3$ & $0.52$ & $0.52$ & $0.083$ \\
\hline
\hline
\end{tabular}
  %}
\end{table}

%
%
% D -> nu nubar
%
Bounds on the combinations of Wilson coefficients in Eq.~\eqref{eq:variables} of  the EFT with LH and RH neutrinos are
obtained from the measurements of Tab.~\ref{tab:ExperimentalInputs} and recasts given  in Eqs.~(\ref{eq:recast_D0_to_pi0}-\ref{eq:darkphoton}).
For each of the bounds we turn on only a single $x_k$. We use the hadronic coefficients $A_{k}^{h_c\to F}$   from Tab.~\ref{tab:A_coeffs}
to obtain limits on the NP  coefficients, presented in Tab.~\ref{tab:xCoeff_limits}.
The bound on  $x_{SP-}$ or  $x_{SP}$ from $D^0\to \textit{invisible}$ agrees with Ref.~\cite{Bause:2020xzj}.
The limit on   $x_{SP}$  from $D^0\to \textit{invisible}$ is almost two  orders of magnitude stronger than from $D^0 \to \pi^0\,\nu\bar\nu$.
The  coefficient $x_{LR-}$ remains weakly constrained as only searches
with narrow $q^2$-windows probe this coupling. Extending the analysis of $D^0\to\omega + \textit{invisible}$ or
$\Lambda_c\to p +\textit{invisible}$ to a bigger $q^2$-window could  improve these bounds.

Besides the general WET limits on $x_k$ presented in Tab.~\ref{tab:xCoeff_limits} additional constraints  exist within  SMEFT  and $\nu$SMEFT.
Upper limits on the  chirality-preserving  four-fermion operators with left-handed neutrinos, and right-handed neutrinos $O_{QN}, O_{uN}$ (\ref{eq:withN}) have been obtained from  large missing transverse energy (MET) plus jet  searches at the LHC \cite{Hiller:2024vtr,Hiller:2025hpf} 
\begin{align}
  x_{LR\pm} & \lesssim 2 \left(\frac{2\pi v^2}{\alpha_e \Lambda^2}\right)^2 |C_{uc}^{4F}|^2 \, , \\ \nonumber
 \left| C_{uc}^{4F} \right|^2 &  =\sum_{ij} \bigg(\left|C_{\ell q}^{(1)ij} +C_{\ell q}^{(3)ij}\right|^2 
 +\left|C^{ij}_{u \ell}\right|^2 \bigg) \\ & + \sum_{pr}\bigg(\left|C^{pr}_{QN}\right|^2+\left|C^{pr}_{uN}\right|^2\bigg)
\end{align}
with  $C_{uc}^{4F}/\Lambda^2  < 0.22/\text{TeV}^2$~ \cite{Hiller:2025hpf}~\footnote{Daniel Wendler, private communication.}.
Here, $ij$ ($pr$) are left-handed (right-handed) neutrino flavors, which are contributing incoherently.
We obtain
\begin{align}
    x_{LR\pm} &\lesssim 270 \quad\quad\text{(MET+jet)} \, . 
  \end{align}
  Due to the incoherent sum, all operators and all flavors are constrained. This feature makes the MET plus jet searches from the LHC and in the future from the high-luminosity LHC (HL-LHC) particularly important
  for invisible decays.
   For lepton universal (LU), charged lepton flavor conserving (cLFC) and general flavor structures
bounds   exploiting the $SU(2)_L$ link
between processes into charged dileptons and into dineutrinos have been obtained from Drell-Yan data \cite{Bause:2020xzj}
\begin{align}
  x_{LR\pm} &\lesssim 68\quad\quad\text{(LU)}\nonumber \, ,  \\
  x_{LR\pm} &\lesssim 392\quad\quad\text{(cLFC)}  \, , \label{eq:SU2L_bounds}\\
  x_{LR\pm} &\lesssim 1432\quad\quad\text{(general)} \, . \nonumber
\end{align}
Since the MET plus jet analysis interpreted in \cite{Hiller:2024vtr,Hiller:2025hpf}   required a full theory simulation, and
in view of the sizable uncertainties we conservatively use for now  (\ref{eq:SU2L_bounds}) to estimate upper SMEFT limits.
The limits from  high-$p_T$  are  considerably stronger then the current limits  on $ x_{LR\pm}$ from  $D$-decays, see Tab.~\ref{tab:xCoeff_limits}.

\subsection{Constraints on ALP parameters \label{sec:alp-param}}

\begin{figure}
  \centering
  \includegraphics[width=0.48\textwidth]{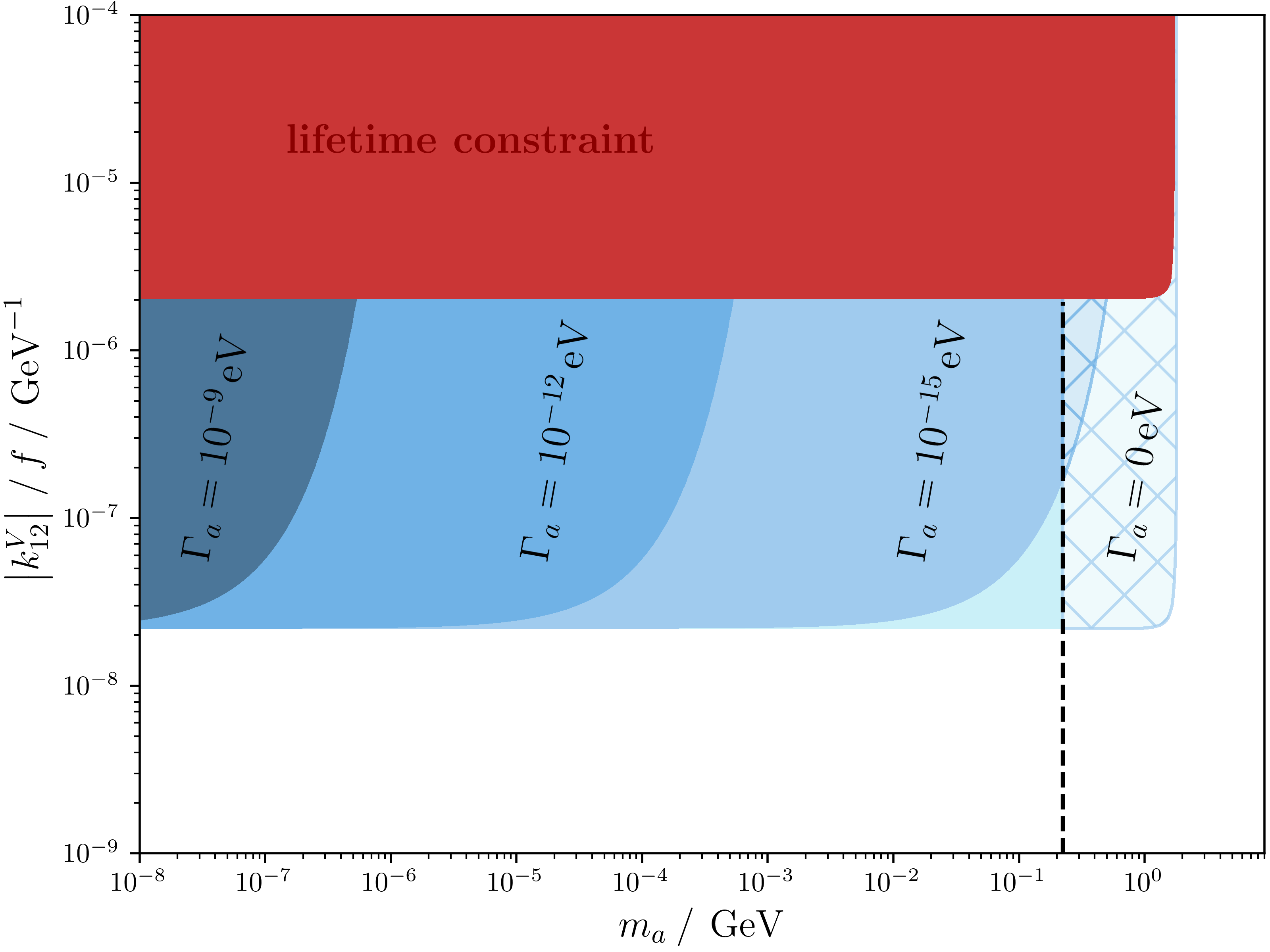}
  \caption{Excluded regions (blue)   of  $\left|k^V_{12}\right|/f$
  depending on the ALP mass $m_a$  for various ALP decay widths $\Gamma_a=\{10^{-9}\eVunit,10^{-12}\eVunit,10^{-15}\eVunit,0\,\eVunit \}$ (from left to right).
  The larger the width, the weaker the constraint.
  The  exclusion region stems from the recast (\ref{eq:cleo-tau}) 
  of  $D^+\to \pi^+ +  invisible$  valid for $m_a<0.224\,\mathrm{GeV}$, the latter indicated by a dashed vertical line. 
   The lighter colored, hatched region to the right of the dashed line is unconstrained. The lifetime constraint (red) of
  $D^+$~\eqref{eq:BRs_unaccounted} is orders of magnitude weaker.
 }
    \label{fig:D_pi_nothing_bound}
\end{figure}

\begin{table}[t]
  \setlength{\tabcolsep}{2pt}
  \centering
  %\captionsetup{format=plain}
  \caption{
    Upper limits on ALP couplings from  charm decays to invisibles  for
  $m_a = 1.2\,\mathrm{GeV}$ and $m_a = 0\,\mathrm{GeV}$, both with $\Gamma_a = 0$, see also Tab.~\ref{tab:xCoeff_limits}. ``n.a.'' indicates that presently no usable data is available, but the decay mode is in principle sensitive to the coupling, as opposed to ``-'', where this is not the case.  For $D^+\to\pi^++\textit{invisible}$
  our results agree with Ref.~\cite{MartinCamalich:2020dfe}. }
  \label{tab:xCoeff_limits_ALPs}
  \centering
  \renewcommand{\arraystretch}{1.5}
  \resizebox{0.49\textwidth}{!}{
  \begin{tabular}{
    l
    c
    c
    c
    c
    c
}
\hline
\hline
&  \multicolumn{2}{c}{$m_a = 1.2\,\mathrm{GeV}$} & \phantom{ab} & \multicolumn{2}{c}{$m_a=0\,\mathrm{GeV}$}\\
\cmidrule[0.5pt](l{0.75em}r{0.75em}){2-3}\cmidrule[0.5pt](l{0.75em}r{0.75em}){5-6}
& $\left|k^V_{12}\right|/f$\hspace{0.5cm} & $\left|k^A_{12}\right|/f$ & & $\left|k^V_{12}\right|/f$\hspace{0.5cm} & $\left|k^A_{12}\right|/f$\\
\cmidrule[0.5pt](l{0.75em}r{0.75em}){2-6}
& \multicolumn{5}{c}{$[10^{-7}\mathrm{GeV}^{-1}]$} \\
\hline
\hline
$D^0\to \pi^0+\textit{invisible}$   & $2.4$          & -         && n.a.     & -    \\
$D^+\to \pi^++\textit{invisible}$   & n.a.           & -         && $0.22$   & -    \\
$D^0 \to \omega+\textit{invisible}$ & -              & n.a.      && -        & $0.70$\\
$\Lambda_c \to p+\textit{invisible}$& n.a.           & n.a.      && $1.3$    & $2.4$\\
\hline
\hline
Best overall & $2.4$ & n.a. && $0.22$ & $0.70$ \\
\hline
\hline
\end{tabular}
  }
\end{table}

We show  in Fig.~\ref{fig:D_pi_nothing_bound} the excluded  region (blue)  of the coupling
$\left|k_{12}^V\right|/f$ and the ALP mass  for various decay widths $\Gamma_a$.
The larger the latter, the weaker the constraints. 
For instance, for $\Gamma = 10^{-12}\eVunit$ we can not probe masses
$m_a \gtrsim 10^{-3}\GeV$   with the invisible decay as the ALP is expected
to mostly decay inside the detector to
$D^+\to \pi^+\gamma\gamma$ or $D^+\to\pi^+ e^+e^-$. 
  ALP constraints from the latter decays are relevant but are beyond the scope of this work.
The recast of  $D^+\to\pi^+ + invisible$  only applies for masses up to  $m_a=0.224\,\mathrm{GeV}$, indicated by the dashed, vertical line.
As they give the strongest constraints, we work out  in the following limits 
for $\Gamma_a=0$.

In Tab.~\ref{tab:xCoeff_limits_ALPs} we present bounds on vector and axial-vector ALP couplings
 for two benchmark  masses $m_a=0 $ and  $m_a=1.2 \, \mathrm{GeV}$, corresponding to the mass windows available by experimental searches, see Sec.~\ref{sec:recast}.
All searches are so far only performed in one of the two mass windows.
Entries with ``-''  indicate that there is no sensitivity to this coupling.
The entries with "n.a.",  short for  "not  available", indicate no data but decays are  in principle probing the coupling.
Specifically the axial-vector coupling $k^A_{12}$ is only constrained for small $m_a$.
We note that the mass benchmarks are representative for their respective mass window as long as they are sufficiently far from the kinematic endpoint of large $m_a$
at which the decay feels phase space suppression and limits weaken.

Additional constraints on the vector ALP couplings can be obtained from $D^0\to \pi^0 a$ via Eq.~\eqref{eq:recast_D0_to_pi0} applicable for  $1.05 < m_a/\,\mathrm{GeV} < 1.38$.
Our results improve on previous works ~\cite{Geng:2022kmf,Beltran:2023nli}  where no recast of the $D^0 \to \pi^0 \nu \bar \nu$ search has been performed and the experimental limit
reported in  Tab.~\ref{tab:ExperimentalInputs} has been used for the whole $m_a$ region  without factoring in a different signal shape.

\subsection{Constraints on   $Z^\prime$ parameters \label{sec:Zp-param}}

\begin{figure}
  \includegraphics[width=0.49\textwidth]{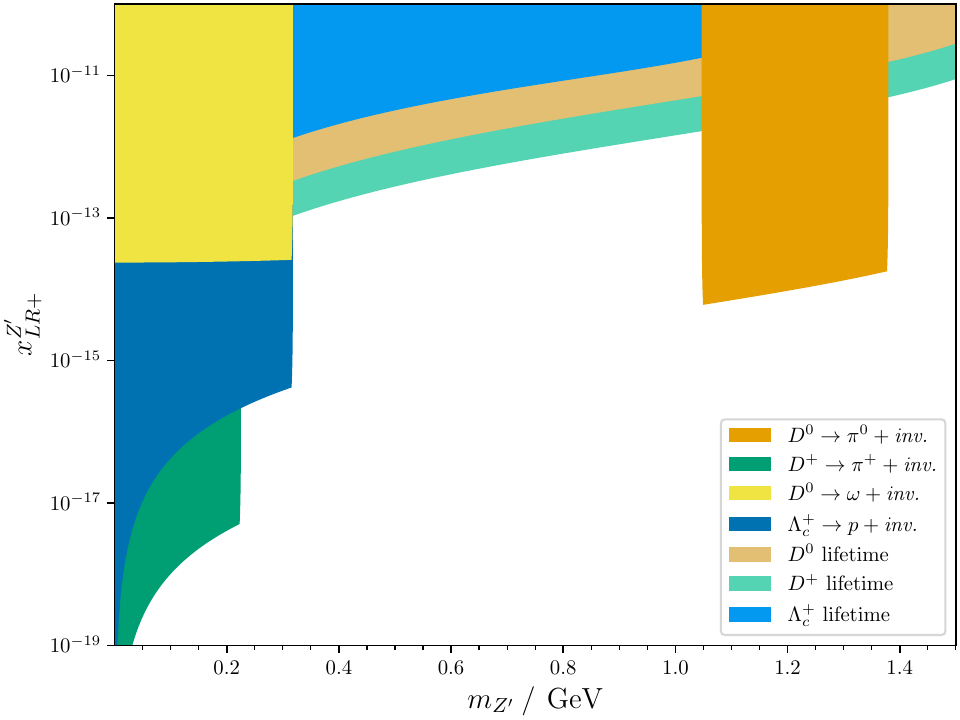}
  \caption{Upper limits on the coefficient $x_{LR}^{Z^\prime}\equiv x_{LR+}^{Z^\prime}=x_{LR-}^{Z^\prime}$~\eqref{eq:variables_Zp} in long-lived $Z^\prime$ models from
  different decay modes, including lifetime constraints~\eqref{eq:BRs_unaccounted}.}
  \label{fig:Zp_longlived_constraints}
\end{figure}

\begin{table}[t]
  \setlength{\tabcolsep}{2pt}
  \centering
  %\captionsetup{format=plain}
  \caption{
    Upper limits on $Z^\prime$ couplings $x_k^{Z^\prime}$  from  Eq.~\eqref{eq:variables_Zp} in the benchmark (\ref{eq:ZpBMV}), see also Tab.~\ref{tab:xCoeff_limits}.
  The columns with $x_{LR}^{Z^\prime}$ are single Wilson coefficient limits, see Eq.~\eqref{eq:variables_SMEFT}.}
  \label{tab:xCoeff_limits_Zp}
  \centering
  \renewcommand{\arraystretch}{1.2}
  %\renewcommand{\arraystretch}{0.8}
  %\resizebox{0.49\textwidth}{!}{
  \begin{tabular}{
    l
    c
    c
    c
    c
    c
    c
    c
}
\hline
\hline
& \multicolumn{5}{c}{BM$_V$ $Z^\prime$} \\
\cmidrule[0.5pt](l{0.75em}r{0.75em}){2-6}
& $x_{LR+}^{Z^\prime}$ & $x_{LR-}^{Z^\prime}$ & $x_{LR}^{Z^\prime}$ & $x_D^{Z^\prime}$ & $x_{D5}^{Z^\prime}$ \\
$h_c\to F$  &  $[10^{-14}]$ & $[10^{-14}]$ & $[10^{-14}]$ & \multicolumn{2}{c}{$[10^{-14}\mathrm{GeV}^2]$}   \\
\hline
\hline
$D^0\to $        & -               & -           & -       & -               & -     \\
\\
$D^0\to \pi^0$   & $2.9$           & -           & $2.9$   & $3.1$           & -     \\
$D^+\to \pi^+$   & $85.9$          & -           & $85.9$  & $5.1\cdot 10^4$ & -     \\
\\
$D^0 \to \omega$ & $2.0\cdot 10^4$ & $471$       & $460$   & $475$           & $444$ \\
\\
$\Lambda_c \to p$& $24.6$          & $287$       & $132$   & $63.3$          & $614$ \\
\hline
\hline
Best overall & $2.9$ & $287$ & $2.9$ & $3.1$ & $444$ \\
\hline
\hline
\end{tabular}
  %}
\end{table}

For $Z^\prime$ models constraints can be separated into the case of a long-lived one
and an invisibly decaying one. For the former the $Z^\prime$ is on-shell and we expect only
events for $q^2=m_{Z^\prime}^2$, while for the latter we have an off-shell $Z^\prime$ and expect a  $q^2$-distribution with shape depending on the model parameters.
A clean boundary between those two is limited by the experimental resolution in $q^2$.

For a long-lived $Z^\prime$, and to a lesser extend also for the invisibly decaying one, the current experimental bounds are limited to
roughly two mass windows, see Sec.~\ref{sec:recast}.
In Fig.~\ref{fig:Zp_longlived_constraints} we present bounds of the coupling $x_{LR}^{Z^\prime}\equiv x_{LR+}^{Z^\prime}=x_{LR-}^{Z^\prime}$~\eqref{eq:variables_Zp}
as a function of the $Z^\prime$ mass.
The mass windows of the existing searches are  visible. The lifetime constraints, see Eq.~\eqref{eq:BRs_unaccounted},
have only outside these regions an impact, the strongest is coming from the $D^+$. The exclusion regions from
$D \to (\pi^0, \omega) +\textit{inv.}$~and $\Lambda_c \to p +\textit{inv.}$~are new results of this work.

For an invisibly decaying $Z^\prime$  using benchmark BM$_V$~\eqref{eq:ZpBMV}, we present bounds on the combinations of couplings, defined in Eq.~\eqref{eq:variables_Zp},
in Tab.~\ref{tab:xCoeff_limits_Zp}. 
As the mass  $m_{Z^\prime}=1 \,$ GeV  is in the large mass search window, and  $\Gamma_{Z^\prime}/m_{Z^\prime}=0.1$,  it follows that couplings, which are not  constrained by  $D^0\to\pi^0+\textit{invisible}$ are only weakly constrained.

\subsection{$D$-$\bar D$-mixing contributions}

We comment briefly on $D$-mixing constraints in the light NP models.
For light ALPs $m_a \approx 0$ they are  obtained as  \cite{MartinCamalich:2020dfe}
\begin{equation}
  \frac{\left|k_{12}^V\right|}{f} \lesssim 8.3 \cdot 10^{-7}\:\mathrm{GeV}^{-1} \, , 
  \quad \frac{\left|k_{12}^A\right|}{f} \lesssim 4.3 \cdot 10^{-7}\:\mathrm{GeV}^{-1} \, .
\end{equation}
Constraints allowing for CP-violation are one order of magnitude stronger, yet,
slightly weaker  than the rare decay limits given in Tab.~\ref{tab:xCoeff_limits_ALPs}.

We naively extrapolate heavy $Z^\prime$ results \cite{Bause:2019vpr,Bause:2022jes} to obtain a rough estimate for the $Z^\prime$-model
\begin{equation} \label{eq:Zpmix}
 \frac{\left|( \mathcal{C}_{L}^{Z^\prime} )^2+ (\mathcal{C}_{R}^{ Z^\prime})^2 -X \mathcal{C}_{L}^{ Z^\prime} \mathcal{C}_{R}^{ Z^\prime}\right|}{m_D^2-m^2_{Z^\prime}} \lesssim 6 \cdot 10^{-13}\: \mathrm{GeV}^{-2}
\end{equation}
yielding
\begin{equation}
  x_{LR\pm}^{Z^\prime}  \lesssim 2 \cdot 10^{-12}
\end{equation}
for $m_{Z^\prime} = 1\,\mathrm{GeV}$, weaker than the rare decay bounds  Tab.~\ref{tab:xCoeff_limits_Zp}. 
The factor $X$  includes ratios of  hadronic matrix elements and RG-running, which we do not consider due  to the proximity of scales, $X \sim 5.1$.
The constraint (\ref{eq:Zpmix}) can be evaded by tuning $\mathcal{C}_L \approx X \mathcal{C}_R$ or  $\mathcal{C}_L \approx X/ \mathcal{C}_R$~\cite{Bause:2019vpr}.

In addition, constraints from the lifetime difference in the $D$-$\bar D$-system on operators with light particles  have been worked out
in \cite{Kumar:2024ivx}. Limits are, depending on the Dirac structure of the operator, either weaker or at most comparable to the ones from rare decays presented in Table.~\ref{tab:xCoeff_limits}.
Exceptions arise for some operators such as tensors in scenarios with finite invisible fermion masses of several hundred MeV, which also suppress phase space in the charm decays.

\section{Predictions}
\label{sec:UpperLimitsBRs}

We summarize   predictions for the $h_c\to F + \textit{invisible}$  branching ratios  in the  EFT, ALP, and $Z^\prime$-models.
We use the constraints on the couplings from Sec.~\ref{sec:EC}.
In Tab.~\ref{tab:BR_limits}  (Tab.~\ref{tab:BR_limits_SMEFT}) we give upper limits on   the  branching ratio for light LH \& RH neutrinos (in the $d=6$ SMEFT), in Tab.~\ref{tab:BR_limits_ALPs} for ALPs and in Tab.~\ref{tab:BR_limits_Zp} for the $Z^\prime$ model.
For each of the limits we assume only a
single combination of Wilson coefficients $x_k$ is turned on while all others are set to zero.

For general patterns  in Tab.~\ref{tab:BR_limits_SMEFT},~\ref{tab:BR_limits},~\ref{tab:BR_limits_ALPs} and \ref{tab:BR_limits_Zp} some comments are in order.
The large differences between  $D^0 \to \pi^0$, $D^+\to \pi^+$, $D^+_s \to K^+$, $\Lambda_c\to p$, $\Xi_c^+\to \Sigma^+$
are caused by the  lifetimes $\tau_{D^0}\simeq \frac{5}{4} \tau_{D_s^+} \simeq \frac{5}{2}\tau_{D^+}$, $\tau_{\Xi_c^+} \simeq 2 \tau_{\Lambda_c}$,
the isospin factor of $2$ for the $\pi^0$ and  to some degree the kinematic cuts of Eq.~\eqref{eq:cuts_charged} for
charged $D_{(s)}$ mesons.
Specifically,  branching ratios induced  by couplings  which have
larger contributions at  low-$q^2$,  see e.g. $x_{LR+}$ (green)  in Fig.~\ref{fig:dBR_dq2_D_to_pinunubar},  are more strongly affected  by these cuts.
Note also the impact of the recast on  the EFT limits given in  Tab.~\ref{tab:BR_limits}. For instance,
the upper limit on  $\mathcal{B}(D^0\to\pi^0\nu\overline{\nu})$ with $x_{SP+}$ turned on is stronger than the experimental value provided in Tab.~\ref{tab:ExperimentalInputs} (which is  based on $x_{LR+}$), since 
the extrapolation outside of the signal region depends on the NP model, for which we correct using Eq.~\eqref{eq:recast_D0_to_pi0}.

Both $Z^\prime$ and ALP constraints are generically dependent on the details of the models and are shown for the $Z^\prime$
benchmark  (\ref{eq:ZpBMV})
and ALP benchmarks $m_a = \{0,1.2\}\GeV$.
The  dependence on $m_a$ is however small if $q^2=m_a^2$ is sufficiently away from the phase space boundary $q^2=(m_{h_c}-m_F)^2$.
For $m_a=1.2$ GeV this holds with the exception of $D^0\to\rho^0/\omega+ a$ because the latter decays are kinematically forbidden.

\begin{figure}
  \centering
  \includegraphics[width=0.48\textwidth]{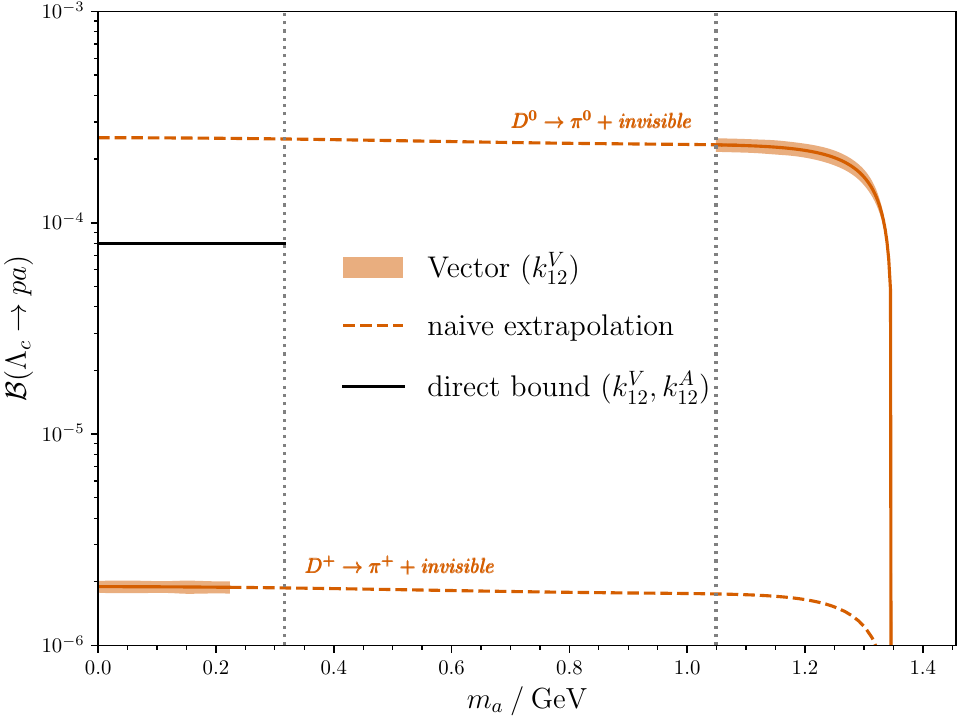}
  \caption{Upper limit on the branching fraction of $\Lambda_c \to p a$ depending on  the ALP mass $m_a$ using 
  $D^{0}\to\pi^{0}+\textit{invisible}$~(\ref{eq:recast_D0_to_pi0})  for large  $m_a$
  and 
  $D^{+}\to\pi^{+}+\textit{invisible}$~(\ref{eq:cleo-tau}) for small  $m_a$,  both valid if only the 
  vector coupling is present. Also shown is the direct bound (black, solid line) $\Lambda_c\to p +\textit{invisible}$~(\ref{eq:darkphoton}) valid for small $m_a$ and both couplings.
   Dashed lines correspond to naive extrapolations  of the bounds outside their respective signal regions. Error bands indicate form factor uncertainties.}
  \label{fig:LambdaC_ALP_Br}
\end{figure}

In Fig.~\ref{fig:LambdaC_ALP_Br} we  show the upper limits on $\mathcal{B}(\Lambda_c\to p a)$ as
a function of the ALP mass using the limits on  $\left|k_{12}^V\right|/f$ from Eqs.~(\ref{eq:recast_D0_to_pi0}),(\ref{eq:darkphoton}) for $\Gamma_a=0$.
Our limits from $D^0\to \pi^0+\textit{invisible}$ agree partially with Ref.~\cite{Geng:2022kmf}
\footnote{Ref.~\cite{Geng:2022kmf} uses  a modified bag model for the form factors
and introduces additional operators $Q_{1(2)} \propto (\bar{u}\,(\gamma_5)\,c)\,a$. However, $Q_{1(2)}$ cannot give
different limits on the branching ratio than 
$Q_{3(4)} \propto (\bar{u}\,\gamma_\mu(\gamma_5)\,c)\,\partial^\mu a$, see Fig.~6 of Ref.~\cite{Geng:2022kmf}, as both sets  are related by e.o.m.
The limits from $Q_{3(4)}$ are of the same magnitude as ours and differences can be accounted for by
the form factors.
}. A similar flatness of the branching ratio 
with respect to $m_a$ can be inferred also from Fig.~\ref{fig:D_ALP_Br_functions} for $D\to \pi a$. Dashed lines indicate a naive extrapolation of limits outside their region of validity (dashed lines).
We find that recasting $D^+\to \tau^+  \nu \to \pi^++\textit{invisible}$ provides presently the strongest limit on $\left|k_{12}^V\right|/f$ \cite{MartinCamalich:2020dfe}, however note that this channel
is eventually limited to $m_a \leq 0.58\,\mathrm{GeV}$  
due to the  2-body decay kinematics  imposed by the $\tau$-resonance \eqref{eq:cuts_charged}.

Models with only scalar- and pseudoscalar contributions  $x_{SP-}\neq 0$ and other $x_k$ vanishing
are most strongly constrained by $D^0\to \textit{invisible}$.
For tensor and dipole couplings $D^0\to \rho^0/\omega + \textit{invisible}$ are golden
modes, with bounds currently limited by the small signal windows of the naive recast.
Future experimental limits on $\Lambda_c\to p + \textit{invisible}$ for a larger  $q^2$ region  would be beneficial as
this mode probes all couplings at the same time.
The same holds  in principle for $D \to \pi \pi + \textit{invisible}$ decays, however
the knowledge of the $D \to \pi \pi $ scalar and tensor form factors should be improved.

\begin{table}[t]
  \setlength{\tabcolsep}{2pt}
  \centering
  %\captionsetup{format=plain}
  \caption{Upper limits on the branching fraction of various $h_c\to F+\textit{invisible}$ decays in SMEFT assuming  lepton universality (LU),
  lepton flavor conservation (cLFC) and general based on Eq.~\eqref{eq:SU2L_bounds}. Limits are in agreement with Ref.~\cite{Bause:2020xzj} with updated meson form factors in our work. $^\ast$ include kinematic cuts, see Eq.~\eqref{eq:cuts_charged}.}
  \label{tab:BR_limits_SMEFT}
  \centering
  \renewcommand{\arraystretch}{1.2}
  \begin{tabular}{
    l
    l
    l
    l
}
\hline
\hline
& \multicolumn{3}{c}{SMEFT} \\
\cmidrule[0.5pt](l{0.75em}r{0.75em}){2-4}
$h_c\to F$  & $\mathcal{B}_{\text{LU}}$ & $\mathcal{B}_{\text{cLFC}}$ & $\mathcal{B}_{\text{general}}$ \\
& $[10^{-7}]$ & $[10^{-6}]$ & $[10^{-5}]$  \\
\hline
\hline
$D^0\to$                & $\phantom{0}0\phantom{.0}$   & $\phantom{0}0\phantom{.0}$         & $0\phantom{.0}$ \\
\\
$D^0\to \pi^0$     & \phantom{2}6.0  & \phantom{2}3.5  & 1.3\\
$D^+\to \pi^+$     & $24.3^\ast$ & $14.0^\ast$ & $5.1^\ast$\\
$D_s^+\to K^+$     & $\phantom{2}5.6^\ast$  & $\phantom{2}3.2^\ast$  & $1.2^\ast$\\
\\
$D^0 \to \omega/\rho^0$  & \phantom{2}6.6    &  \phantom{2}3.8    & 1.4  \\
\\
$\Lambda_c\to p$   & 18.8 & 10.8 & 4.0\\
$\Xi_c\to \Sigma^+$& 35.6 & 20.5 & 7.5\\
\\
$D^0\to \pi^+\pi^-$& \phantom{2}4.4    & \phantom{2}2.4    & 0.9  \\
$D^+\to \pi^+\pi^0$& $16.3^\ast$    & $\phantom{2}9.4^\ast$    & $3.4^\ast$  \\
\hline
\hline
\end{tabular}
\end{table}
\begin{table}[t]
  \setlength{\tabcolsep}{2pt}
  \centering
  %\captionsetup{format=plain}
  \caption{Upper limits on the branching ratios of $h_c\to F + \textit{invisible}$ from
  bounds on the Wilson coefficients of LH \& RH neutrino models from Tab.~\ref{tab:xCoeff_limits} for  coupling combinations  $x_k$  switched on.  The values in brackets correspond to the directly measured experimental limits.  ``n.a.'' indicates
  for $D\to\pi\pi$ that no bound is available because of insufficient information on the form factors, see text for details.
   $^\ast$ include kinematic cuts, see Eq.~\eqref{eq:cuts_charged}. The columns with $x_{SP}$ and $x_{LR}$ are single SMEFT coefficient limits, see Eq.~\eqref{eq:variables_SMEFT}.
  }
  \label{tab:BR_limits}
  \centering
  \renewcommand{\arraystretch}{1.2}
  %\resizebox{0.49\textwidth}{!}{
  \begin{tabular}{
    l
    l
    l
    l
    l
    l
    l
    l
}
\hline
\hline
& \multicolumn{7}{c}{Light LH \& RH neutrino}  \\
\cmidrule[0.5pt](l{0.75em}r{0.75em}){2-8}
  & $x_{SP-}$ & $x_{SP+}$ & $x_{SP}$ & $x_{LR-}$ & $x_{LR+}$ & $x_{LR}$ & $x_{T}$  \\
$h_c\to F$ & \multicolumn{7}{c}{Upper limit on branching ratio $\mathcal{B} \: /\: 10^{-4}$} \\
\hline
\hline
$D^0\to$                & $(0.94)$ & $0$         & $(0.94)$    & $0$        & $0$         & $0$        & $0$        \\
\\
$D^0\to \pi^0$          & $0$      & $1.8$       & $0.025$     & $0$        & $0.50$      & $0.50$     & $1.4$      \\
$D^+\to \pi^+$          & $0$      & $8.9^\ast$  & $0.13^\ast$ & $0$        & $2.0^\ast$  & $2.0^\ast$ & $6.9^\ast$ \\
$D_s^+\to K^+$          & $0$      & $1.9^\ast$  & $0.027^\ast$& $0$        & $0.45^\ast$ & $0.45^\ast$& $1.0^\ast$ \\
\\
$D^0 \to \rho^0/\omega$ & $0.0017$ & $0$         & $0.0017$    & $0.85$     & $0.031$     & $0.51$     & $20.1$     \\
\\
$\Lambda_c\to p$        & $0.0056$ & $0.87$      & $0.017$     & $1.6$      & $0.53$      & $1.4$      & $7.6$      \\
$\Xi_c\to \Sigma^+$     & $0.0091$ & $1.5$       & $0.030$     & $3.2$      & $0.93$      & $2.7$      & $14.7$     \\
\\
$D^0 \to \pi^+ \pi^-$   & n.a.     & $0$         & n.a.        & $0.53$     & $0.060$     & $0.35$     & n.a.       \\
$D^+ \to \pi^+ \pi^0$   & n.a.     & $0$         & n.a.        & $1.8^\ast$ & $0.25^\ast$ & $1.3^\ast$ & n.a.       \\
\hline
\hline
\end{tabular}
  %}
\end{table}

\begin{table}[t]
  \setlength{\tabcolsep}{2pt}
  \centering
  %\captionsetup{format=plain}
  \caption{Upper limits on the branching ratios of $h_c\to F + \textit{invisible}$ in   ALPs models
  using the limits 
    from Tab.~\ref{tab:xCoeff_limits_ALPs} for the benchmarks $m_a=1.2\,\mathrm{GeV}$ and $m_a=0\,\mathrm{GeV}$, both with $\Gamma_a = 0$. Values in parentheses are input for the predictions given in this table.
  }
  \label{tab:BR_limits_ALPs}
  \centering
  %\renewcommand{\arraystretch}{0.8}
  %\resizebox{0.49\textwidth}{!}{
  \begin{tabular}{
    l
    l
    l
    l
    l
    l
}
\hline
\hline
&  \multicolumn{2}{c}{$m_a = 1.2\,\mathrm{GeV}$} & \phantom{ab} & \multicolumn{2}{c}{$m_a=0\,\mathrm{GeV}$}\\
\cmidrule[0.5pt](l{0.75em}r{0.75em}){2-3}\cmidrule[0.5pt](l{0.75em}r{0.75em}){5-6}
& $k^V_{12}$\hspace{0.5cm} & $k^A_{12}$ & & $k^V_{12}$\hspace{0.5cm} & $k^A_{12}$\\
$h_c\to F$ & \multicolumn{5}{c}{Branching ratio $\mathcal{B} \: /\: 10^{-4}$} \\
\hline
\hline
$D^0\to \pi^0$          & $(0.58)$    & $0$ && $0.019$   & $0$    \\
$D^+\to \pi^+$          & $2.9$       & $0$ && $(0.096)$ & $0$    \\
$D_s^+\to K^+$          & $1.2$       & $0$ && $0.046$   & $0$    \\
\\
$D^0 \to \rho^0/\omega$\hspace{0.1cm} & $0$       & n.a. && $0$     & $0.039$\\
\\
$\Lambda_c\to p$        & $0.61$      & n.a.&& $0.024$ & $(0.80)$ \\
$\Xi_c\to \Sigma^+$     & $1.1$       & n.a.&& $0.054$ & $1.8$  \\
\hline
\hline
\end{tabular}
  %}
\end{table}

\begin{table}[t]
  \setlength{\tabcolsep}{2pt}
  \centering
  %\captionsetup{format=plain}
  \caption{Upper limits on the branching ratios of $h_c\to F + \textit{invisible}$ in   $Z^\prime$ models  for  the benchmark (\ref{eq:ZpBMV})
  using bounds on the  $x_k^{Z^\prime}$ in Tab.~\ref{tab:xCoeff_limits_Zp}.
  For charm baryons, the branching ratio limits from $x_{D5}^{Z^\prime}$ are weaker than lifetime constraints~\eqref{eq:BRs_unaccounted}.
  }
  \label{tab:BR_limits_Zp}
  \centering
  \renewcommand{\arraystretch}{1.2}
  %\resizebox{0.49\textwidth}{!}{
  \begin{tabular}{
    l
    l
    l
    l
    l
}
\hline
\hline
& \multicolumn{4}{c}{BM$_V$ $Z^\prime$}  \\
\cmidrule[0.5pt](l{0.75em}r{0.75em}){2-5}
  &  $x_{LR-}^{Z^\prime}$ & $x_{LR+}^{Z^\prime}$ & $x_{D}^{Z^\prime}$ & $x_{D5}^{Z^\prime}$ \\
$h_c\to F$ & \multicolumn{4}{c}{Branching ratio $\mathcal{B} \: /\: 10^{-4}$} \\
\hline
\hline
$D^0\to$                & $0$   & $0$         & $0$        & $0$    \\
\\
$D^0\to \pi^0$          & $0$   & $2.9$       & $1.9$      & $0$    \\
$D^+\to \pi^+$          & $0$   & $14.7^\ast$ & $9.8^\ast$ & $0$    \\
$D_s^+\to K^+$          & $0$   & $5.5^\ast$  & $2.5^\ast$ & $0$    \\
\\
$D^0 \to \rho^0/\omega$ & $24$  & $0.014$     & $0.59$     & $378$  \\
\\
$\Lambda_c\to p$        & $763$ & $4.5$       & $16.7$     & $4.3\cdot 10^3$ \\
$\Xi_c\to \Sigma^+$     & $1.7\cdot 10^3$& $7.4$       & $28.6$     & $9.2\cdot 10^3$ \\
\\
$D^0 \to \pi^+ \pi^-$   & $86$     & $0.23$           & n.a.       & n.a.     \\
$D^+ \to \pi^+ \pi^0$   & $373$     & $0.97$          & n.a.       & n.a.     \\
\hline
\hline
\end{tabular}
  %}
\end{table}

\begin{figure}[t]
  \centering
  \includegraphics[width=0.48\textwidth]{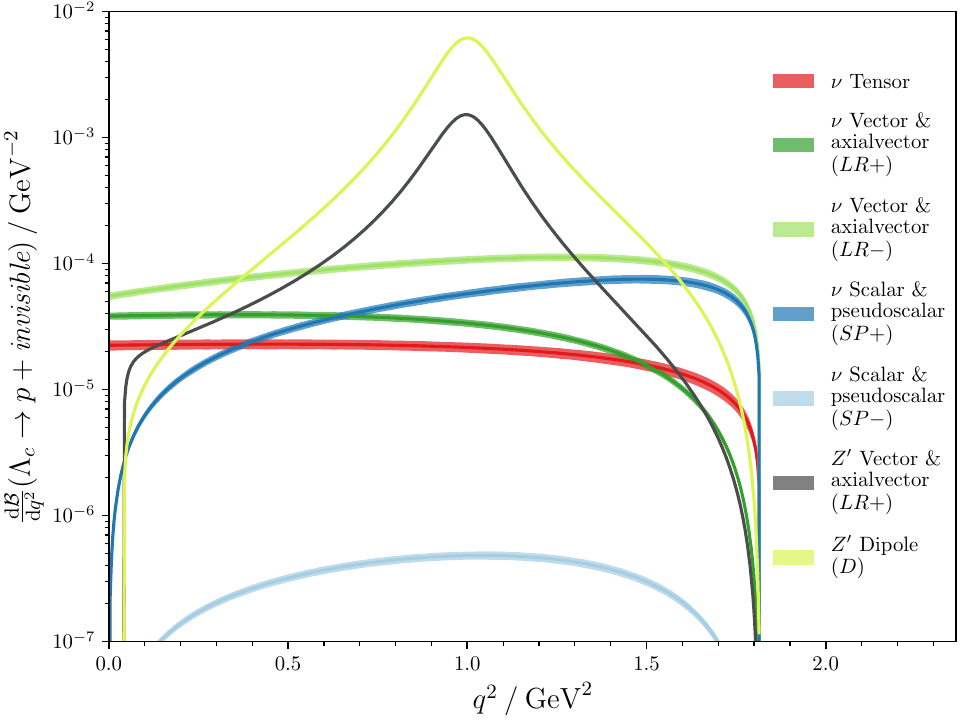}
  \caption{Maximal achievable differential branching fraction of $\Lambda_c \to p + \textit{invisible}$ in EFT (red,green,light green,blue,light blue) and light $Z^\prime$ models (gray, yellow) in BM$_V$~\eqref{eq:ZpBMV}.
  }
  \label{fig:BRLimitsBaryon}
\end{figure}

Furthermore, differential branching ratios can  distinguish different NP models and couplings.
In Fig.~\ref{fig:BRLimitsBaryon}  we  show  the maximum achievable $dB/dq^2(\Lambda_c \to p +inv.)$ in different models
using the constraints from Sec.~\ref{sec:EC}. Apart from the height, the shapes of the $q^2$-distribution are vastly different.
We emphazise the region of  $q^2 \approx 0$, in which most  distributions vanish, except for the tensor (red) and the vector ones (green, lighter green), see also 
 Fig.~\ref{fig:dBR_dq2_D_to_pinunubar} - Fig.~\ref{fig:DiffPlot1}.
In addition, the resonance structure from the $Z^\prime$ with finite width is a smoking gun for this model. 
More general, light LH \& RH neutrinos allow for finite contributions at the kinematic endpoint $q^2=0$ for vector-, axial-vector- or tensor-couplings for baryon decays 
 and vector-, axial-vector-couplings for mesons. In scenarios with other couplings the rate dies off towards the endpoint with 
significant slopes for low-$q^2$, which are also distinguishable. The behavior towards $q^2 \to 0$ is diagnostic:
if observed, for instance,  in $\Lambda_c$ decays this would require scalar, pseudoscalar couplings, which
indicate very specifically the  non-standard phenomena LNV or light sterile neutrinos. In general the whole distribution contains information.
Bounds on $D^0\to\pi^0+\textit{invisible}$ for example in the high-$q^2$ are most constraining  for $x_{SP+}$, while for $x_{LR+}$ this happens at low-$q^2$.
These features can  be exploited experimentally if the $q^2$-binning is  sufficient.
%\clearpage
\section{Conclusions \label{sec:con}}

We  study rare decays of charmed hadrons into invisible final states, which arise 
in extensions of the SM with light and heavy degrees of freedom:
dineutrinos in four-fermion $SU(2)_L \times U(1)_Y$ invariant SMEFT and $\nu$SMEFT operators  induced by heavy BSM mediators, left- and right- handed light neutrinos in four-fermion WET operators from weak-scale NP,
 a light  $Z^\prime$-boson which is long-lived or decays into light and dark fermions,
and an ALP that is sufficiently long-lived to escape the detector undecayed.

We work out achievable branching ratios of two-, three- and four-body decays.
We  find that branching ratios reach $10^{-4}-10^{-3}$ with  chirality-flipping interactions, see Tab.~\ref{tab:BR_limits}, Tab.~\ref{tab:BR_limits_ALPs} and Tab.~\ref{tab:BR_limits_Zp}.
Branching ratios are more constrained in the chirality-preserving  SMEFT,  $\lesssim \text{few} \times 10^{-5}$, see 
Tab.~\ref{tab:BR_limits_SMEFT}, due to the constraints from processes involving charged leptons. Lepton flavor structure  plays a key  role in the link between neutrinos and charged leptons,
making upper limits flavor structure dependent~\cite{Bause:2020xzj}.
In the future MET plus jet searches in $pp$-collisions  will become important \cite{Hiller:2024vtr,Hiller:2025hpf}.

The $d=7$ LNV SMEFT and the  $d=6$ $\nu$SMEFT induce chirality-flipping operators. They are presently best probed with
$D^0 \to invisibles$, reaching scales as low as  $\Lambda_{\text{LNV}} \gtrsim 1.5 $ TeV and $\Lambda_{\nu\text{SMEFT}} \gtrsim 2.1 $ TeV,
allowing for NP with invisibles  in $|\Delta c|=|\Delta u|=1$ transitions to  be just  around the corner.

Not all couplings  with RH neutrinos and in the light mediator models, $Z^\prime$ and ALPs, are presently experimentally constrained, see Tab.~\ref{tab:xCoeff_limits_ALPs}.
For instance, for  larger ALP masses the axial-vector coupling is not probed, leaving  branching ratios of decays $D \to V+invisible$, with vectors $V=\omega, \rho,..$ and  baryons $\Lambda_c \to p+ invisible$
essentially unconstrained,  up to  lifetime constraints (\ref{eq:BRs_unaccounted}), which are weak, at best at the level of $\mathcal{O}(0.1)$.
This highlights the importance of searches in modes beyond $D \to \pi +invisible$ due to the sensitivity to different and additional  couplings.

We perform a recast of the searches in Tab.~\ref{tab:ExperimentalInputs}, see Sec.~\ref{sec:recast}, which results in inproved constraints on ALP models specifically from the 
$D^0 \to \pi^0 \nu \bar \nu$ search.
The methodology could be improved in the future and we encourage the experimental collaborations to broaden their signal regions, and present results in a way that can cleanly be reinterpreted, as
also suggested for the $B \to K^{(*)} \nu \bar \nu$ analysis~\cite{Gartner:2024muk}.

 Correlations enable the identification of  the underlying type of NP, by  comparing shapes of missing energy distributions from different couplings,
 Figs.~\ref{fig:dBR_dq2_D_to_pinunubar} -~\ref{fig:DiffPlot1}, \ref{fig:BRLimitsBaryon} and by comparing 
  branching ratios of different decay modes, Fig.~\ref{fig:correlation_ratio_ALP_Zp}.
 Sharper interpretations of the null test observables analyzed in this work need improved hadronic transition form factors of charmed hadrons, such as from lattice QCD.
Especially limited information exists  presently for  $D \to V$ and $D \to \pi \pi$.
 
The missing energy modes are well-suited for the experiments  Belle II \cite{Belle-II:2018jsg}, BESIII \cite{BESIII:2020nme}, and future $e^+ e^-$-colliders, such as a super tau-charm factory (STCF)
 \cite{Achasov:2023gey} or the Tera-$Z$ facilities FCC-ee \cite{FCC:2018byv,FCC:2025lpp} and CEPC \cite{Ai:2024nmn},
with sizable charm  rates~\cite{Bause:2020xzj,DiCanto:2025fpk}. Since any observation of $c \to u + invisibles$-transitions heralds NP, with sizable branching ratios in decays of $D$-mesons and charm baryons, systematic experimental searches are encouraged.

\acknowledgments

This work is supported by the \textit{Bundesministerium f\"ur Forschung, Technologie und Raumfahrt } -- BMFTR.
We are grateful to Andreas J\"uttner, Yotam Soreq  and Daniel Wendler for useful discussions.
We thank Mustafa Tabet for useful exchanges on the recasts of experimental data.

\section*{Data availability}

The data that support the findings of this article are
openly available~\cite{RawData}.

\appendix

\section{Form factors}\label{sec:FF}

We provide details on the hadronic transition form factors used in our analysis.

\subsection{\texorpdfstring{$D \to \pi$}{D -> pi} form factors}\label{sec:FFDtopi}
\begin{figure}
  \centering
  \includegraphics[width=0.48\textwidth]{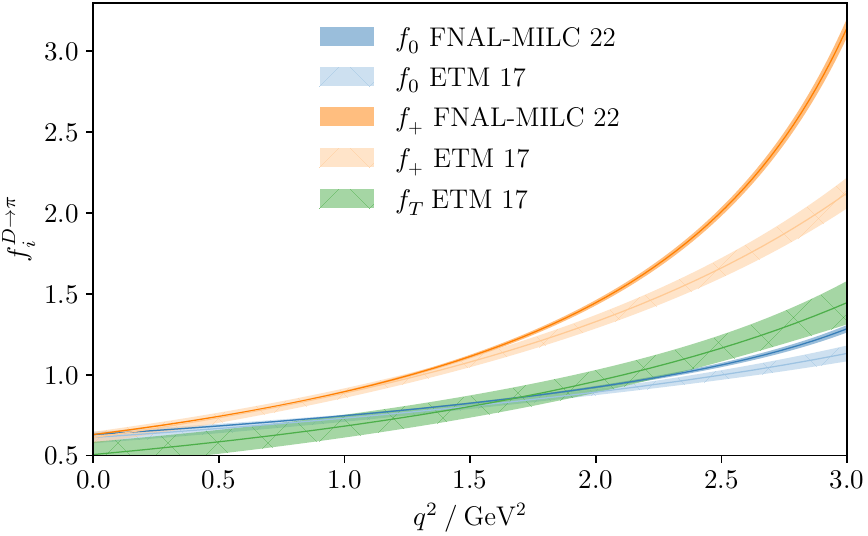}
  \caption{Mesonic form factors $f_{+,0,T}$ of $D\to\pi$ are shown as functions of  $q^2$. The form factors are taken from the ETM collaboration (lighter colors)
  \cite{Lubicz:2017syv,Lubicz:2018rfs} and from the Fermilab Lattice and MILC collaborations \cite{FermilabLattice:2022gku} (darker colors). }
  \label{fig:FormFactor_D_to_pi}
\end{figure}
For $D \to \pi$ transitions only three independent form factors with vector, scalar or tensor currents exist.
They can be parametrized as~\cite{Lubicz:2017syv} 
\begin{equation}
  \begin{aligned}
   & \mel{\pi(p_P)}{\:\bar{u}\gamma_\mu c\:}{D(p_D)} \\
      &\qquad= (p_D + p_P)_\mu f_+(q^2) + (p_D - p_P)_\mu f_-(q^2) \:,
  \end{aligned}
\end{equation}
where $q$ is the momentum transfer given by $q=p_D - p_P$ with $p_D (p_P)$ the momenta of the $D$ meson (pion).
The scalar form factor $f_0$ can be related through the
equations of motion and is given by
\begin{equation}
  \begin{aligned}
    &\mel{\pi(p_P)}{\:\bar{u} c\:}{D(p_D)} \\
    &\qquad = \frac{m_D^2 - m_P^2}{m_c - m_u}\underbrace{\left( f_+(q^2) + \frac{q^2}{m_D^2 - m_P^2} f_-(q^2)\right)}_{ \textstyle \equiv f_0(q^2)} \:.
  \end{aligned}
\end{equation}
The form factors fulfill the endpoint relation $f_+(0)=f_0(0)$.
The tensor form factor is parametrized as \cite{Lubicz:2018rfs} 
\begin{equation}
  \begin{aligned}
    &\mel{\pi(p_P)}{\:\bar{u}\, \sigma^{\mu\nu}\, c\:}{D(p_D)} \\
    &\qquad = \frac{2}{m_D + m_P}\left( p_P^\mu p_D^\nu - p_P^\nu p_D^\mu \right) f_T(q^2) \:.
  \end{aligned}
\end{equation}
The form factors from lattice QCD are shown in Fig.~\ref{fig:FormFactor_D_to_pi},
from the
ETM collaboration \cite{Lubicz:2017syv,Lubicz:2018rfs} (lighter colors) and for $f_{+,0}^{D\to \pi}$ also from the
Fermilab Lattice and MILC collaborations \cite{FermilabLattice:2022gku} (darker colors).
Both agree for lower $q^2$ but differ  at  higher $q^2$ as already noted in Ref.~\cite{FermilabLattice:2022gku}.
For $D_s\to K$ we use the 
same form factors as for $D\to\pi$, as supported by \cite{FermilabLattice:2022gku}.

\subsection{\texorpdfstring{$D \to V$}{D -> V} form factors}\label{sec:FFDtoV}
For $D\to V$ transitions there are seven independent form factors, which we define~\cite{Lin:2025cmn,Melikhov:2000yu,Khodjamirian:2020btr}
\begin{equation}
    \begin{aligned}
        &\mel{V(p,\eta^\ast)}{\bar{u} \gamma_5  c}{D(p_D)} = -\frac{m_V(\eta^\ast \cdot q)}{m_c+m_u}  A_0(q^2) \:,  \\
        &\mel{V(p,\eta^\ast)}{\bar{u} \gamma^\mu  c}{D(p_D)} = i \frac{2 V(q^2)}{m_D+m_V} \epsilon^{\mu\nu\rho\sigma} \eta^{\ast}_\nu p_{D\,\rho} p_{\sigma} \:,\\
        &\mel{V(p,\eta^\ast)}{\bar{u} \gamma^\mu \gamma_5  c}{D(p_D)} \\
        &\quad\quad\quad= 2 m_V A_0(q^2) \frac{\eta^\ast \cdot q}{q^2} q^\mu \\
        &\quad\quad\quad\quad+ (m_D+m_V) A_1(q^2) \left(\eta^{\ast\,\mu} - \frac{\eta^\ast \cdot q}{q^2} q^\mu\right) \\
        &\quad\quad\quad\quad - A_2(q^2) \frac{\eta^\ast \cdot q}{m_D+m_V} \left(p_D^\mu + p^\mu - \frac{m_D^2 -m_V^2}{q^2} q^\mu\right) \:,\\
        &\mel{V(p,\eta^\ast)}{\bar{u}\sigma_{\alpha \beta} \gamma_5 c}{D(p_D)} \\
        &\quad\quad\quad= \left(\eta^\ast_\alpha (p_D+p)_\beta - \eta^\ast_\beta (p_D+p)_\alpha\right) T_1(q^2) \\
        &\quad\quad\quad\quad+ \left(\eta^\ast_\alpha q_\beta - \eta^\ast_\beta q_\alpha\right) \frac{m_D^2 - m_V^2}{q^2} \left( T_2(q^2) - T_1(q^2)\right) \\
        &\quad\quad\quad\quad+ (\eta^\ast \cdot q) \left(q_\alpha (p_D+p)_\beta - q_\beta (p_D+p)_\alpha\right) \\
        &\quad\quad\quad\quad\quad\times\left(\frac{T_3(q^2)}{m_D^2-m_V^2}  + \frac{T_2(q^2)-T_1(q^2)}{q^2} \right) \:,\\
    \end{aligned}
\end{equation}
using the convention $\epsilon_{0123} = +1$. It is common to define two additional form factors via the expressions
\begin{equation}
    \begin{aligned}
        A_{12}(q^2) &\equiv -\frac{1}{16 m_D m_V^2 (m_D+m_V)} \bigg( \lambda(q^2,m_D^2,m_V^2)A_2(q^2) \\
        &\quad - (m_D+m_V)^2(m_D^2 - m_V^2 -q^2)A_1(q^2) \bigg)  \:,\\
        T_{23}(q^2) &\equiv -\frac{1}{8 m_D m_V^2 (m_D-m_V)} \bigg( \lambda(q^2,m_D^2,m_V^2)T_3(q^2) \\
        &\quad- (m_D^2-m_V^2)(m_D^2 + 3 m_V^2 -q^2)T_2(q^2) \bigg) \:.
    \end{aligned}
\end{equation}
For our numerical calculation we use the LCSR determination of the vector and axial-vector form factors from
Ref.~\cite{Lin:2025cmn} and the quark-model calculation from Ref.~\cite{Melikhov:2000yu} for the tensor form factors.
In the heavy quark limit tensor and vector factors are related, see Ref.~\cite{Burdman:2000ku}.

\subsection{\texorpdfstring{$\Lambda_c \to p$}{Lambda\textunderscore c -> p} form factors}\label{sec:FFLambdaC}
The 10 independent form
factors for $\Lambda_c \to p$ decays read in  the helicity-based definition \cite{Feldmann:2011xf}, 
\begin{equation}
  \begin{aligned}
      &\mel{p(p_p,\lambda_p)}{\:\bar{u}\: c\:}{\Lambda_c(p_{\Lambda_c},\lambda_{\Lambda_c})} \\
      &\qquad= f_0(q^2)\frac{m_{\Lambda_c}-m_p}{m_c-m_u}\bar{u}_p(p_p,\lambda_p) u_{\Lambda_c}(p_{\Lambda_c},\lambda_{\Lambda_c}) \:,
    \end{aligned}
    \end{equation}
    \begin{equation}
      \begin{aligned}
      &\mel{p(p_p,\lambda_p)}{\:\bar{u}\gamma_5 c\:}{\Lambda_c(p_{\Lambda_c},\lambda_{\Lambda_c})} \\
      &\qquad= g_0(q^2)\frac{m_{\Lambda_c}+m_p}{m_c+m_u}\bar{u}_p(p_p,\lambda_p) \gamma_5 u_{\Lambda_c}(p_{\Lambda_c},\lambda_{\Lambda_c}) \:,
      \end{aligned}
      \end{equation}
\begin{equation}
\begin{aligned}
    &\mel{p(p_p,\lambda_p)}{\:\bar{u}\gamma^\mu c\:}{\Lambda_c(p_{\Lambda_c},\lambda_{\Lambda_c})} \\
    &\quad= \bar{u}_p(p_p,\lambda_p)  \left[ f_0(q^2)(m_{\Lambda_c} - m_p ) \frac{q^\mu}{q^2}  \right. \\
            &\qquad\quad+ f_+(q^2)\frac{m_{\Lambda_c}+m_p}{s_+}\left( p_{\Lambda_c}^\mu + p_p^\mu - (m_{\Lambda_c}^2-m_p^2) \frac{q^\mu}{q^2}  \right) \\
            &\qquad\quad+ \left. f_{\bot}(q^2) \left( \gamma^\mu - \frac{2m_p}{s_+}p_{\Lambda_c}^\mu - \frac{2m_{\Lambda_c}}{s_+}p_p^\mu \right) \right] \\
            &\qquad \times u_{\Lambda_c}(p_{\Lambda_c},\lambda_{\Lambda_c}) \:,
\end{aligned}
\end{equation}
\begin{equation}
\begin{aligned}
    &\mel{p(p_p,\lambda_p)}{\:\bar{u}\gamma^\mu\gamma_5 c\:}{\Lambda_c(p_{\Lambda_c},\lambda_{\Lambda_c})} \\
    &\qquad= - \bar{u}_p(p_p,\lambda_p) \gamma_5 \left[ g_0(q^2)(m_{\Lambda_c} + m_p ) \frac{q^\mu}{q^2}  \right. \\
            &\qquad\quad+ g_+(q^2)\frac{m_{\Lambda_c}-m_p}{s_-}\left( p_{\Lambda_c}^\mu + p_p^\mu - (m_{\Lambda_c}^2-m_p^2) \frac{q^\mu}{q^2}  \right) \\
            &\qquad\quad+ \left. g_{\bot}(q^2) \left( \gamma^\mu + \frac{2m_p}{s_-}p_{\Lambda_c}^\mu - \frac{2m_{\Lambda_c}}{s_-}p_p^\mu \right) \right] \\
            &\qquad \times u_{\Lambda_c}(p_{\Lambda_c},\lambda_{\Lambda_c}) \:,\\
\end{aligned}
\end{equation}
\begin{equation}
\begin{aligned}
        &\mel{p(p_p,\lambda_p)}{\:\bar{u}i\sigma^{\mu\nu} c\:}{\Lambda_c(p_{\Lambda_c},\lambda_{\Lambda_c})} \\
        &\quad= \bar{u}_p(p_p,\lambda_p)  \Bigg[ 2h_+(q^2) \frac{p_{\Lambda_c}^\mu p_p^\nu  - p_{\Lambda_c}^\nu p_p^\mu }{s_+}  \\
            &\qquad+ h_{\bot}(q^2) \left( \frac{m_{\Lambda_c}+m_p}{q^2} \left(q^\mu \gamma^\nu - q^\nu \gamma^\mu\right) \right. \\
            &\qquad\quad- \left. 2\left(\frac{1}{q^2} + \frac{1}{s_+}\right)  \left( p_{\Lambda_c}^\mu p_p^\nu  - p_{\Lambda_c}^\nu p_p^\mu \right) \right) \\
            &\qquad+ \tilde{h}_{+}(q^2) \left(i \sigma^{\mu\nu}- \frac{2}{s_-} \left[ m_{\Lambda_c} \left( p_p^\mu\gamma^\nu - p_p^\nu\gamma^\mu \right)  \right. \right. \\
            &\qquad\quad- \left. \left. m_p \left( p_{\Lambda_c}^\mu \gamma^\nu - p_{\Lambda_c}^\nu \gamma^\mu \right) + \left( p_{\Lambda_c}^\mu p_p^\nu  - p_{\Lambda_c}^\nu p_p^\mu \right) \right] \right) \\
            &\qquad+ \tilde{h}_{\bot}(q^2) \frac{m_{\Lambda_c}-m_p}{q^2s_-} \bigg( \\
            &\qquad\quad \left(m_{\Lambda_c}^2 - m_p^2 - q^2 \right)\left( \gamma^\mu p_{\Lambda_c}^\nu - \gamma^\nu p_{\Lambda_c}^\mu \right) \\
            &\qquad\quad- \left( m_{\Lambda_c}^2 - m_p^2 + q^2\right)\left(\gamma^\mu p_{p}^\nu - \gamma^\nu p_{p}^\mu \right)  \\
            &\qquad\quad+ 2 \left(m_{\Lambda_c}-m_p\right) \left( p_{\Lambda_c}^\mu p_p^\nu  - p_{\Lambda_c}^\nu p_p^\mu \right)  \bigg) \Bigg] \\
            &\qquad \times u_{\Lambda_c}(p_{\Lambda_c},\lambda_{\Lambda_c}) \:.
\end{aligned}
\end{equation}
The tensor matrix element is inferred from the dipole definition in Ref.~\cite{Feldmann:2011xf} through the
relation $\sigma^{\mu\nu}\gamma_5 = \frac{i}{2} \epsilon^{\mu\nu\alpha\beta}\sigma_{\alpha\beta}$.
The following endpoint relations hold \cite{Golz:2021imq}
\begin{equation}
  \begin{aligned}
    f_0(0) &= f_+(0)\,, \quad \quad &g_\perp(q^2_{\text{max}}) &= g_+(q^2_{\text{max}})\,, \\
    g_0(0) &= g_+(0)\,, \quad \quad &\tilde{h}_\perp(q^2_{\text{max}}) &= \tilde{h}_+(q^2_{\text{max}})\,, \\
    h_\perp(0) &= \tilde{h}_\perp(0)\:. \quad\quad &
  \end{aligned}
\end{equation}
We take the form factors $f_0,g_0,f_+,g+,f_{\bot},g_{\bot},h_+,h_{\bot},\tilde{h}_+$ 
and $\tilde{h}_{\bot}$ from  lattice QCD computations~\cite{Meinel:2017ggx}.
For  $\Xi_c^+\to\Sigma^+$ we use the ones from  $\Lambda_c\to p$, which are related in the  flavor symmetry limit \cite{Bause:2020xzj}.

\subsection{\texorpdfstring{$D \to \pi \pi$}{D -> pipi} form factors}\label{sec:FFDtoPiPi}
There are seven independent transversity form factors $\mathcal{F}_i(q^2,p^2,P\cdot q)$ for $D^0\to \pi^+\pi^-$ transitions depending
on the momentum transfer $q^\mu = p_D^\mu - p^\mu$, the momentum of the dipion system $p^\mu = p_1^\mu + p_2^\mu$ and the angle $\theta_{\pi^+}$
between the $\pi^+$ momentum and the negative direction of flight of the $D$-meson in the dipion-cms. The angle is related to the scalar product $P\cdot q$, where $P^\mu = p_1^\mu - p_2^\mu$.
Here $p_1(p_2)$ are the momenta of the $\pi^+(\pi^-)$ and $p_D$ the momentum of the $D$-meson.

We define the form factors as ~\cite{Faller:2013dwa,Descotes-Genon:2019bud}
\begin{equation}
\begin{aligned}
    -i &\mel{\pi^+(p_1)\pi^-(p_2)}{\bar{u}\gamma_\mu \gamma_5 c}{D(p_D)} \\
    &\quad\quad\quad= \frac{2}{\mathcal{N}_{nr}^{\pi\pi}\sqrt{\lambda_D}} \left(p^\mu - \frac{q\cdot p}{q^2} q^\mu\right) \mathcal{F}_0 \nonumber\\
    &\quad\quad\quad\quad+\frac{\sqrt{p^2}}{\mathcal{N}_{nr}^{\pi\pi}\sqrt{q^2 \lambda_P}}\bigg(P^\mu - \frac{4(q\cdot p)(q\cdot P)}{\lambda_D} p^\mu \\
    &\quad\quad\quad\quad\quad\quad\quad+ \frac{4p^2 (q\cdot P)}{\lambda_D}q^\mu\bigg) \mathcal{F}_{\parallel} \nonumber\\
    &\quad\quad\quad\quad+ \frac{q^\mu}{\mathcal{N}_{nr}^{\pi\pi}\sqrt{q^2}^3} \mathcal{F}_t \nonumber  \, , \\
    i &\mel{\pi^+(p_1)\pi^-(p_2)}{\bar{u}\gamma_\mu  c}{D(p_D)} \\
    &\quad\quad\quad= + \frac{4\sqrt{p^2}}{\mathcal{N}_{nr}^{\pi\pi}\sqrt{q^2 \lambda_D \lambda_P}} i \epsilon_{\mu\alpha\beta\gamma} q^\alpha p_1^\beta p_2^\gamma \mathcal{F}_\perp  \, ,
\end{aligned}
\end{equation}
\\
\begin{equation}
\begin{aligned}
    &\mel{\pi^+(p_1)\pi^-(p_2)}{\bar{u}\sigma^{\mu\nu}\gamma_5 c}{D(p_D)} \\
    &\quad\quad\quad= \frac{2(p^{\mu} q^\nu - q^\mu p^\nu)}{\mathcal{N}_{nr}^{\pi\pi}\sqrt{q^2\lambda_D}} \bigg(\mathcal{F}_0^T \\
    &\quad\quad\quad\quad\quad\quad\quad+ \frac{(P\cdot q)\sqrt{p^2}}{\sqrt{q^2\lambda_P}}\left(\mathcal{F}_\perp^T - 2\frac{p\cdot q}{\sqrt{\lambda_D}} \mathcal{F}_\parallel^T\right) \bigg)  \\
    &\quad\quad\quad\quad-\frac{2(p^{\mu} P^\nu - P^\mu p^\nu)\sqrt{p^2}}{\mathcal{N}_{nr}^{\pi\pi} \sqrt{\lambda_D\lambda_P}} \mathcal{F}^T_{\perp}\\
    &\quad\quad\quad\quad+ \frac{(P^{\mu} q^\nu - q^\mu P^\nu)\sqrt{p^2}}{\mathcal{N}_{nr}^{\pi\pi}q^2\sqrt{\lambda_P}}\left(\mathcal{F}_{\parallel}^T - 2 \frac{p\cdot q}{\sqrt{\lambda_D}} \mathcal{F}_\perp^T \right)  \, ,
\end{aligned}\label{eq:FFs_DtoPiPi}
\end{equation}
\begin{equation}
\begin{aligned}
    -i &\mel{\pi^+(p_1)\pi^-(p_2)}{\bar{u} \gamma_5 c}{D(p_D)} = \frac{1}{\mathcal{N}_{nr}^{\pi\pi} \sqrt{q^2}}\frac{1}{m_c+m_u}\mathcal{F}_t  \, ,\nonumber
\end{aligned}
\end{equation}
with normalization
\begin{equation}
    \quad\mathcal{N}_{nr}^{\pi\pi} = \frac{G_F \alpha_e}{2^7 \pi^4 m_D} \sqrt{\pi\frac{\sqrt{\lambda_D \lambda_P}}{m_D p^2}}\:,
\end{equation}
where $\lambda_P = \lambda(p^2,m_\pi^2,m_\pi^2)$ and  $\epsilon_{0123} = +1$. Our definition slightly differs
from Ref.~\cite{Faller:2013dwa,Descotes-Genon:2019bud} in the normalization, such
that we agree with the notation in~\cite{Bause:2020xzj,DiCanto:2025fpk}.
For brevity we omitted here the kinematic dependence of the form factors, e.g. $\mathcal{F}_i \equiv \mathcal{F}_i(q^2,p^2,P\cdot q)$. Form factors of other matrix elements can be obtained via e.o.m. or the relation $\sigma_{\mu\nu} = - \frac{i}{2} \epsilon_{\mu\nu\alpha\beta} \sigma^{\alpha\beta} \gamma_5$.
For our numerical analysis we use the form factors $\mathcal{F}_{0,\perp,\parallel}$ of Ref.~\cite{DiCanto:2025fpk} obtained in a data-driven approach from isospin related $D^+\to\pi^+\pi^-e^+\nu_e$ measurements.
We refrain from evaluating expressions with other form factors e.g. $\mathcal{F}_t, \mathcal{F}_i^T$ as no data-driven approach is available for those and other methods show some disagreement, see ~\cite{DiCanto:2025fpk}.

For the form factors of $D^-\to \pi^-\pi^0$ we follow a similar approach as the data-driven approach in Ref.~\cite{DiCanto:2025fpk}.
However in this case the isospin related decay is $D^0\to \pi^-\pi^0 e^+\nu_e$ and only the contribution from the $\rho$ meson contributes.
We fix the  normalization by the measurement $\mathcal{B}(D^0\to \pi^-\pi^0 e^+\nu_e) = (1.439\pm 0.033 \pm 0.027)\cdot 10^{-3}$ \cite{BESIII:2024lxg} and
use $r_V=V(0)/A_1(0)=1.548\pm 0.079 \pm 0.041$ and $r_2=A_2(0)/A_1(0)=0.823\pm 0.056\pm 0.026$~\cite{BESIII:2024lxg} instead of the values in Ref.~\cite{DiCanto:2025fpk} for this decay only.

%--------+---------+---------+---------+---------+---------+---------+---------+

\end{document}